\documentclass[sigconf,screen]{acmart}

\usepackage{url}
\usepackage[inline]{enumitem}
\usepackage{xcolor}
\usepackage{graphicx}
\usepackage{multirow}
\usepackage{microtype}
\usepackage{flushend}

\newcommand{\zc}[1]{\textcolor{red}{\textbf{ZC}: #1}}
\newcommand{\mulong}[1]{\textcolor{orange}{#1}}

\AtBeginDocument{%
  \providecommand\BibTeX{{%
    \normalfont B\kern-0.5em{\scshape i\kern-0.25em b}\kern-0.8em\TeX}}}

\setcopyright{rightsretained}
\acmPrice{}
\acmDOI{10.1145/3540250.3549138}
\acmYear{2022}
\copyrightyear{2022}
\acmSubmissionID{fse22main-p648-p}
\acmISBN{978-1-4503-9413-0/22/11}
\acmConference[ESEC/FSE '22]{Proceedings of the 30th ACM Joint European Software Engineering Conference and Symposium on the Foundations of Software Engineering}{November 14--18, 2022}{Singapore, Singapore}
\acmBooktitle{Proceedings of the 30th ACM Joint European Software Engineering Conference and Symposium on the Foundations of Software Engineering (ESEC/FSE '22), November 14--18, 2022, Singapore, Singapore}

\begin{document}

\title{Psychologically-Inspired, Unsupervised Inference of Perceptual Groups of GUI Widgets from GUI Images}

\author{Mulong Xie}
\email{mulong.xie@anu.edu.au}
\affiliation{
  \institution{Australian National University}
  \country{Australia}
}

\author{Zhenchang Xing}
\email{zhenchang.xing@anu.edu.au}
\affiliation{
  \institution{CSIRO's Data61 \& ANU}
  \country{Australia}
}

\author{Sidong Feng}
\email{sidong.feng@monash.edu}
\affiliation{
  \institution{Monash University}
  \country{Australia}
}

\author{Xiwei Xu}
\email{xiwei.xu@data61.csiro.au}
\affiliation{
  \institution{CSIRO's Data61}
  \country{Australia}
}

\author{Liming Zhu}
\email{liming.zhu@data61.csiro.au}
\affiliation{
  \institution{CSIRO's Data61 \& UNSW}
  \country{Australia}
}

\author{Chunyang Chen}
\email{chunyang.chen@monash.edu}
\affiliation{
  \institution{Monash University}
  \country{Australia}
}

\renewcommand{\shortauthors}{Mulong Xie, Zhenchang Xing, Sidong Feng, Xiwei Xu, Liming Zhu, Chunyang Chen}

\begin{abstract}

    Graphical User Interface (GUI) is not merely a collection of individual and unrelated widgets, but rather partitions discrete widgets into groups by various visual cues, thus forming higher-order perceptual units such as tab, menu, card or list.
    The ability to automatically segment a GUI into perceptual groups of widgets constitutes a fundamental component of visual intelligence to automate GUI design, implementation and automation tasks.
    Although humans can partition a GUI into meaningful perceptual groups of widgets in a highly reliable way, perceptual grouping is still an open challenge for computational approaches.
    Existing methods rely on ad-hoc heuristics or supervised machine learning that is dependent on specific GUI implementations and runtime information.
    Research in psychology and biological vision has formulated a set of principles (i.e., Gestalt theory of perception) that describe how humans group elements in visual scenes based on visual cues like connectivity, similarity, proximity and continuity.
    These principles are domain-independent and have been widely adopted by practitioners to structure content on GUIs to improve aesthetic pleasantness and usability.
    Inspired by these principles, we present a novel unsupervised image-based method for inferring perceptual groups of GUI widgets.
    Our method requires only GUI pixel images, is independent of GUI implementation, and does not require any training data.
    The evaluation on a dataset of 1,091 GUIs collected from 772 mobile apps and 20 UI design mockups shows that our method significantly outperforms the state-of-the-art ad-hoc heuristics-based baseline.
    Our perceptual grouping method creates opportunities for improving UI-related software engineering tasks.



\end{abstract}



\maketitle

\section{introduction}

\begin{figure}
    \centering
    \includegraphics[width=0.47\textwidth]{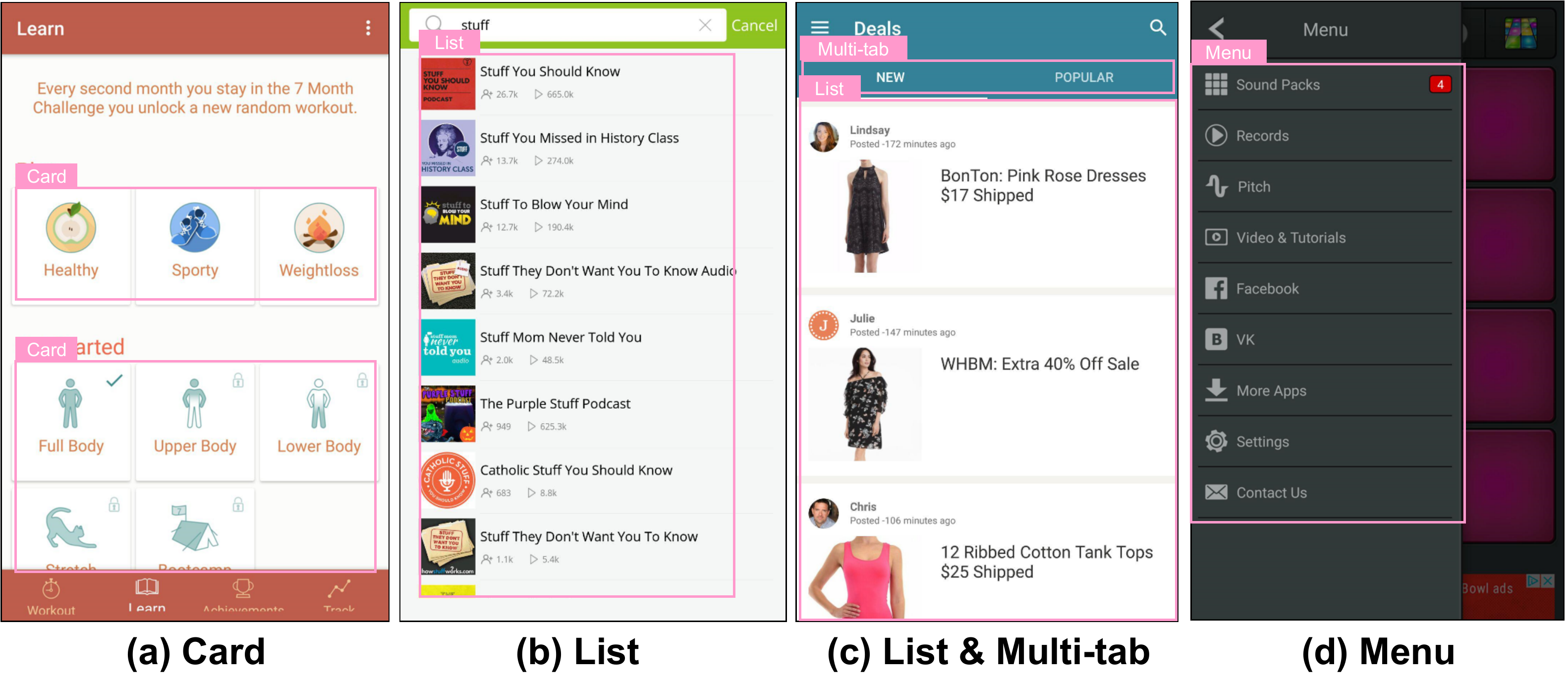}
    \caption{Examples of perceptual groups of GUI widgets (perceptual groups are highlighted in pink box in this paper)}
    \label{fig:examples}
    \vspace{-1mm}
\end{figure}

We do not just see a collection of separated texts, images, buttons, etc., on GUIs.
Instead, we see perceptual groups of GUI widgets, such as card, list, tab and menu shown in Figure~\ref{fig:examples}.
Forming perceptual groups is an essential step towards visual intelligence.
For example, it helps us decide which actions are most applicable to certain GUI parts, such as, clicking a navigation tab, expanding a card, scroll the list.
This would enable more efficient automatic GUI testing~\cite{ase2019humanoid, issta2019learninguielementinteractions}.
As another example, screen readers~\cite{appleaccess, googleaccess} help visually impaired users access applications by reading out content on GUI.
Recognizing perceptual groups would allow screen readers to navigate the GUI at higher-order perceptual units (e.g., sections) efficiently~\cite{screenrecognition}.
Last but not least, GUI requirements, designs and implementations are much more volatile than business logic and functional algorithms.
With perceptual grouping, modular, reusable GUI code can be automatically generated from GUI design images, which would expedite rapid GUI prototyping and evolution~\cite{remaui,redraw}.  

\begin{figure}[t]
    \centering
    \includegraphics[width=0.4\textwidth]{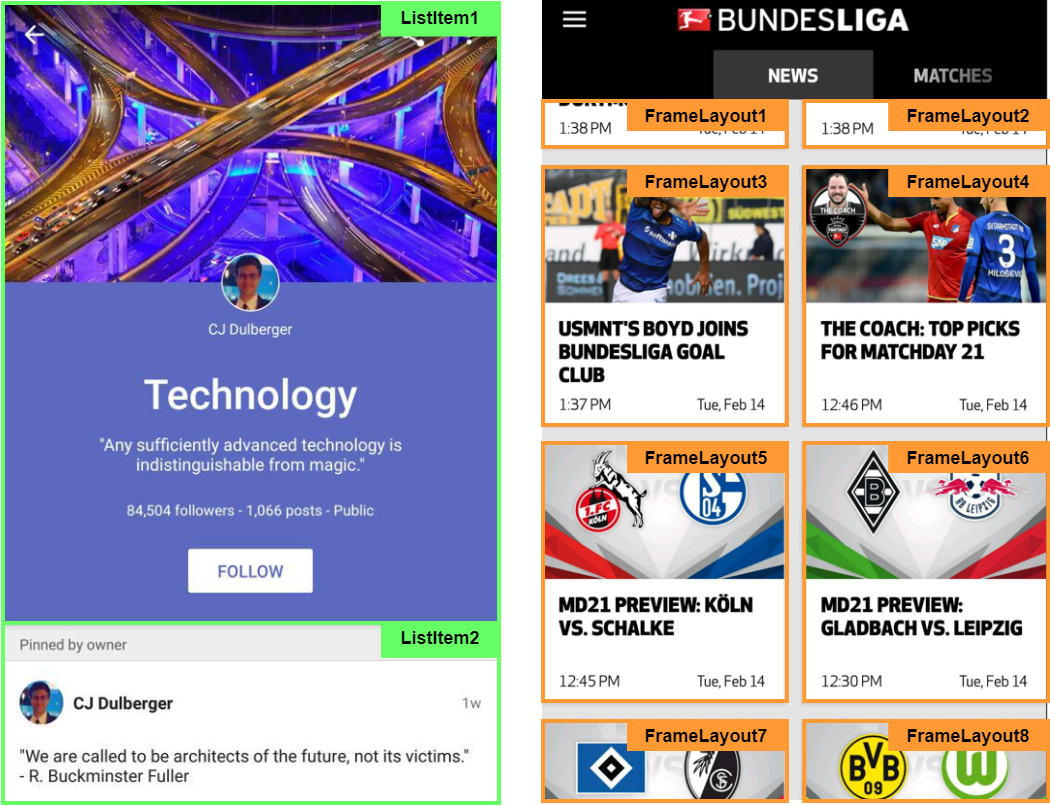}
    \caption{Implemented view hierarchy does not necessarily correspond to perceptual groups}
    \label{fig:rico-wrong}
    \vspace{-4mm}
\end{figure}

Although humans can intuitively see perceptual groups of GUI widgets, current computational approaches are limited in partitioning a GUI into meaningful groups of widgets.
Some recent work~\cite{beltramelli2018pix2code, chunyangui2code} relies on supervised deep learning methods (e.g., image captioning~\cite{imgcaption1, deeplearning_Nature}) to generate a view hierarchy for a GUI image.
This type of method is heavily dependent on GUI data availability and quality.
To obtain sufficient GUI data for model training, they use GUI screenshots and view hierarchies obtained at application runtime.
A critical quality issue of such runtime GUI data is that runtime view hierarchies often do not correspond to intuitive perceptual groups due to many implementation-level tricks.
For example, in the left GUI in Figure~\ref{fig:rico-wrong}, the two ListItems in a ListView has no visual similarity (a large image versus some texts), so they do not form a perceptual group.
In the right GUI, a grid of cards form a perceptual group but is implemented as individual FrameLayouts.
Such inconsistencies between the implemented widget groups and the human's perceptual groups make the trained models unreliable to detect perceptual groups of GUI widgets.

Decades of psychology and biological vision research have formulated the Gestalt theory of perception that explains how humans see the whole rather than individual and unrelated parts.
It includes a set of principles of grouping, among which \textit{connectedness}, \textit{similarity}, \textit{proximity} and \textit{continuity} are the most essential ones \cite{CognitivePsychology, gestaltpsychology}.
Although these principles and other related UI design principles such as CRAP~\cite{crap} greatly influence how designers and developers structure GUI widgets~\cite{gelstal_for_ui_design}, they have never been systematically used to automatically infer perceptual groups from GUI images.
Rather, current approaches~\cite{remaui, screenrecognition} rely on ad-hoc and case-specific rules and thus are hard to generalize on diverse GUI designs. 

\begin{figure*}[t]
    \centering
    \includegraphics[width=0.98\textwidth]{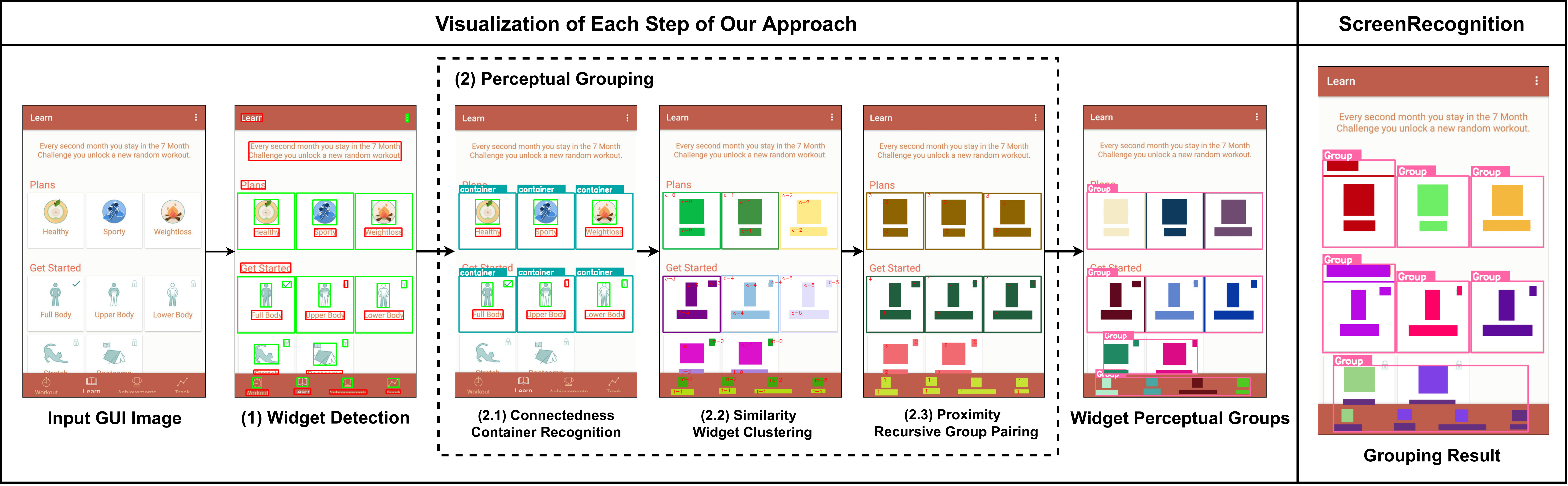}
    \caption{Left: Our approach overview: (1) Enhanced UIED~\cite{UIED-full-fse} for GUI widget detection; (2) Gestalt-principles inspired perceptual grouping. Right: Grouping result of the state of of the art heuristic-based approach ScreenRecognition~\cite{screenrecognition}}
    \vspace{-2mm}
    \label{fig:overallarch}
\end{figure*}

In this work, we systematically explore the Gestalt principles of grouping and design the first psychologically-inspired method for visual inference of perceptual groups of GUI widgets.
Our method requires only GUI pixel images and is independent of GUI implementations.
Our method is unsupervised, thus removing the dependence on problematic GUI runtime data.
As shwon in Figure~\ref{fig:overallarch}, our method enhances the state-of-the-art GUI widget detection method (UIED~\cite{UIED-full-fse, xie2020uied}) to detect elementary GUI widgets.
Following the Gestalt principles, the method first detects containers (e.g., card, list item) with complex widgets by the connectedness principle.
It then clusters visually similar texts (or non-text widgets) by the similarity principle and further groups clusters of widgets by the proximity principles.
Finally, based on the widget clusters, our method corrects erroneously detected or missing GUI widgets by the continuity principle (not illustrated in Figure~\ref{fig:overallarch}, but can be seen in Figure~\ref{fig:misdetection}).

At the right end of Figure~\ref{fig:overallarch}, we show the grouping result by the state-of-the-art heuristic-based method Screen Recognition~\cite{screenrecognition}.
Screen Recognition incorrectly partitions many widgets into groups, such as the bottom navigation bar and the four widgets above the bar, the card on the left and the text above the card.
It also fails to detect higher-order perceptual groups, such as groups of cards.
In contrast, our approach correctly recognizes the bottom navigation bar and the top and middle row of cards.
Although the text label above the left card is very close to the card, our approach correctly recognizes the text labels as separate widgets rather than as a part of the left card.
Our approach does not recognize the two cards just above the bottom navigation bar because these two cards are partially occluded by the bottom bar.
However, it correctly recognizes the two blocks of image and text and detects them as a group.
Clearly,  the grouping results by our approach correspond more intuitively to human perception than those by Screen Recognition.

For the evaluation, we construct two datasets: one contains 1,091 GUI screenshots from 772 Android apps, and the other contains 20 UI prototypes from a popular design tool Figma~\cite{figma}.
To ensure the validity of ground-truth widget groups, we manually check all these GUIs and confirm that none of these GUIs has the perception-implementation misalignment issues shown in Figure~\ref{fig:rico-wrong}.
We first examine our enhanced version of UIED and observe that the enhanced version reaches a 0.626 F1 score for GUI widget detection, which is much higher than the original version (0.524 F1).
With the detected GUI widgets, our perceptual grouping approach achieves the F1-score of 0.593 on the 1,091 app GUI screenshots and 0.783 F1-score on the 20 UI design prototypes.
To understand the impact of GUI widget misdetections on perceptual grouping, we extract the GUI widgets directly from the Android app's runtime metadata (i.e., ground-truth widgets) and use the ground-truth widgets as the inputs to perceptual grouping.
With such ``perfectly-detected'' GUI widgets, our grouping approach achieves a 0.672 F1-score on app GUIs.
In contrast, Screen Recognition~\cite{screenrecognition} performs very poorly: 0.123 F1 on the ground-truth widgets and 0.092 F1 on the detected widgets for app screenshots, and 0.232 F1 on the detected widgets for UI design prototypes.
Although our grouping results sometimes do not exactly match the ground-truth groups, our analysis shows that some of our grouping results still comply with how humans perceive the widget groups because there can be diverse ways to structure GUI widgets in some cases.

To summarize, this paper makes the following contributions:
\begin{itemize}
	\item A robust, psychologically-inspired, unsupervised visual inference method for detecting perceptual groups of GUI widgets on GUI images, the code is released on GitHub\footnote{https://github.com/MulongXie/GUI-Perceptual-Grouping}.
	\item A comprehensive evaluation of the proposed approach and the in-depth analysis of the performance with examples.
	\item An analysis of how our perceptual grouping method can improve UI-related SE tasks, such as UI design, implementation and automation.
\end{itemize}

\section{GUI Widget Detection}
\label{detection}

Our approach is a pixel-only approach.
It does not assume any GUI metadata or GUI implementation about GUI widgets. 
Instead, our approach detects text and non-text GUI widgets directly from GUI images. 
To obtain the widget information from pixels, it enhances the state-of-the-art GUI widget detection technique UIED~\cite{xie2020uied}.
In order to fit with subsequent perceptual grouping, our approach mitigates the incorrect merging of GUI widgets in the containers by UIED and simplifies the widget classification of UIED.

\subsection{UIED-Based GUI Widget Detection}
\label{sec:widgetdetection}
UIED comprises three steps: (1) GUI text detection, (2) non-text GUI widget detection and (3) merging of text and non-text widgets.
UIED uses an off-the-shelf scene text detector EAST~\cite{east} to identify text regions in the GUI images.
EAST is designed for handling nature scene images that differ from GUI images, such as figure-background complexity and lighting effects.
Although EAST outperforms traditional OCR tool Tesseract~\cite{tesseract-ocr}, we find the latest OCR tool developed by Google~\cite{googleocr} achieves the highest accuracy of GUI text recognition (see Section \ref{sec:detectioneval}).
Therefore, in our use of UIED, we replace EAST with the Google OCR tool.

For locating non-text widgets, our approach adopts the design of UIED that uses a series of traditional, unsupervised image processing algorithms rather than deep-learning models (e.g., Faster RCNN~\cite{frcnn} or YOLO~\cite{yolo}).
This design removes the data dependence on GUI implementation or runtime information while accurately detecting GUI widgets.
UIED then merges the text and non-text detection results.
The purpose of this merging step is not only to integrate the identified GUI widgets but also to cross-check the results.
Because non-text widget detection inevitably extracts some text regions, UIED counts on the OCR results to remove these false-positive non-text widgets.
Specifically, this step checks the bounding box of all candidate non-text widget regions and removes those intersected with text regions resulting from the OCR.

\subsection{Improvement and Simplification of UIED}
We find that the UIED detection results often miss some widgets in a container (e.g. card).
The reason is that, in order to filter out invalid non-text regions and mitigate over-segmentation that wrongly segments a GUI widget into several parts, UIED checks the widgets' bounding boxes and merges all intersected widgets regions into a big widget region.
This operation may cause the wrong neglection of valid GUI widgets that are enclosed in some containers.
Therefore, we equip our GUI widget detector with a container recognition algorithm (see Section~\ref{containerrecog}) to mitigate the issue.
If a widget is recognized as a container, then all its contained widgets are kept and regarded as proper GUI widgets rather than noises. 

The original UIED classifies non-text GUI widgets as specific widget categories (e.g., image, button, checkbox).
In contrast, our GUI widget detector only distinguishes text from non-text widgets.
Although GUI involves many types of non-text widgets, there is no need of distinguishing actual classes of non-text widgets for perceptual grouping.  
GUI widget classes indicate different user interactions and GUI functionalities, but widgets with different classes can form a perceptual group as long as they have similar visual properties, such as size, shape, relative position and alignment with other widgets.
Therefore, we do not distinguish different classes of non-text widgets.
However, we need to distinguish non-text widgets from text widgets, as they have very different visual properties and need to be clustered by different strategies (see Section~\ref{fig:overallarch}).

\section{GUI Widget Perceptual Grouping}
After obtaining text and non-text widgets on a GUI image, the next step is to partition GUI widgets into perceptual groups (or blocks of items) according to their visual and perceptual properties.

\subsection{Gestalt Laws and Approach Overview}
Our approach is inspired by psychology and biological-vision research. 
Perceptual grouping is a cognitive process in which our minds leap from comprehending all of the objects as individuals to recognizing visual patterns through grouping visually related elements as a whole ~\cite{FundamentalsofCognition}.
This process affects the way we design GUI layouts~\cite{crap} from alignment, spacing and grouping tool support~\cite{figma,sketch} to UI design templates~\cite{dribbble} and GUI frameworks~\cite{wix}
It also explains how we perceive GUI layouts.
For instance, in the examples in Figure~\ref{fig:examples}, we subconsciously observe that some visually similar widgets are placed in a spatially similar way and identify them as in a group (e.g. a card, list, multitab or menu).

Previous studies rely on ad-hoc, rigid heuristics to infer UI structure without a systematic theoretical foundation. 
Our approach is the first attempt to tackle the perceptual grouping of GUI widgets guided by an influential psychological theory (named Gestalt psychology \cite{gestaltpsychology}) that explains how the human brain perceives objects and patterns.
Gestalt psychology's core proposition is that human understands external stimuli as wholes rather than as the sums of their parts \cite{TheGestaltApproachtoChange}.
Based on the proposition, the Gestaltists studied perceptual grouping~\cite{FundamentalsofCognition} systematically and summarized a set of ``gestalt laws of grouping''~\cite{CognitivePsychology}. 
In our work, we adopt the four most effective principles which greatly influence GUI design ~\cite{gelstal_for_ui_design} in practice as the guideline for our approach design:  
\begin{enumerate*}
    \item connectedness
    \item similarity
    \item proximity and 
    \item continuity.
\end{enumerate*}

We define a group of related GUI widgets as a \textit{layout block} of \textit{items}.
A typical example is a \textit{list} of \textit{list items} in the GUI, or a \textit{card} displaying an image and some texts.
The fundamental intuition is: if a set of widgets have similar visual properties and are placed in alignment with similar space between each other, they will be ``perceived'' as in the same \textit{layout block} by our approach according to the Gestalt principles.
In detail, our approach consists of four grouping steps in accordance with four Gestalt principles.
First, it identifies containers along with their contained widgets that fulfil the connectedness law. 
Second, it uses an unsupervised clustering method DBSCAN~\cite{dbscan} to cluster text or non-text GUI widgets based on their spatial and size similarities.
Next, it groups proximate and spatially aligned clusters to form a larger layout block following the proximity law.
Finally, in line with the continuity principle, our approach corrects some mistakes of GUI widget detection by checking the consistency of the groups' compositions.


\subsection{Connectedness - Container Recognition}
\label{containerrecog}
In Gestalt psychology, the principle of uniform connectedness is the strongest principle concerned with relatedness~\cite{andyrutledge_3laws}. 
It implies that we perceive elements connected by uniform visual properties as being more related than those not connected. 
The forms of the connection can be either a line connecting several elements or a shape boundary that encloses a group of related elements.
In the GUI, the presentation of the connectedness is usually a box container that contains multiple widgets within it, and all the enclosed widgets are perceived as in the same group.
Thus, the first step of our grouping approach is to recognize the containers in a GUI.

In particular, we observe that a container is visually a (round) rectangular wireframe enclosing several children widgets.
The card is a typical example of such containers, as shown in Figure~\ref{fig:examples}(a).
Therefore, with the detected non-text widgets, the algorithm first checks if a widget is of rectangle shape by counting how many straight lines its boundary comprises and how they compose.
Specifically, we apply the geometrical rule that a rectangle's sides are made of 4 straight lines perpendicular to each other.
Subsequently, our approach determines if the widget's boundary is a wireframe border by checking if it is connected with any widgets inside its boundary. 
If a widget satisfies the above criteria, it will be identified as a container, and all widgets contained within it are partitioned into the same perceptual group.

\subsection{Similarity - Widget Clustering}
\label{sec:widgetclustering}

The principle of similarity suggests that elements are perceptually grouped together if they are similar to each other~\cite{wikipedia_gestalt}.
Generally, similarity can be observed in aspects of various visual cues, such as size, color, shape or position. 
For example, in the second GUI of Figure~\ref{fig:examples}, the image widgets are of the same size and aligned with each other in the same way (i.e., same direction and spacing), so we visually perceive them as a group.
Likewise, the text pieces on the right of the image widgets are perceptually similar even though they have different font styles and lengths because they have the same alignment with neighbouring texts.

\subsubsection{Initial Widget Clustering}
Our approach identifies similar GUI widgets by their spatial and visual properties and aggregates similar GUI widgets into blocks by similarity-based clustering.
It clusters texts and non-text widgets through different strategies.
In general, similar non-text widgets in the same block (e.g. a \textit{list}) usually have similar sizes and align to one another vertically or horizontally with the same spacing.
Texts in the same block are always left-justified or top-justified (assume left-to-right text orientation), but their sizes and shapes can vary significantly because of different lengths of text contents.
Thus, the approach clusters the non-text widgets by their center points $(Center_X, Center_Y)$ and areas, and it clusters texts by their top-left corner $(Top, Left)$


\begin{figure}[t]
    \centering
    \includegraphics[width=0.47\textwidth]{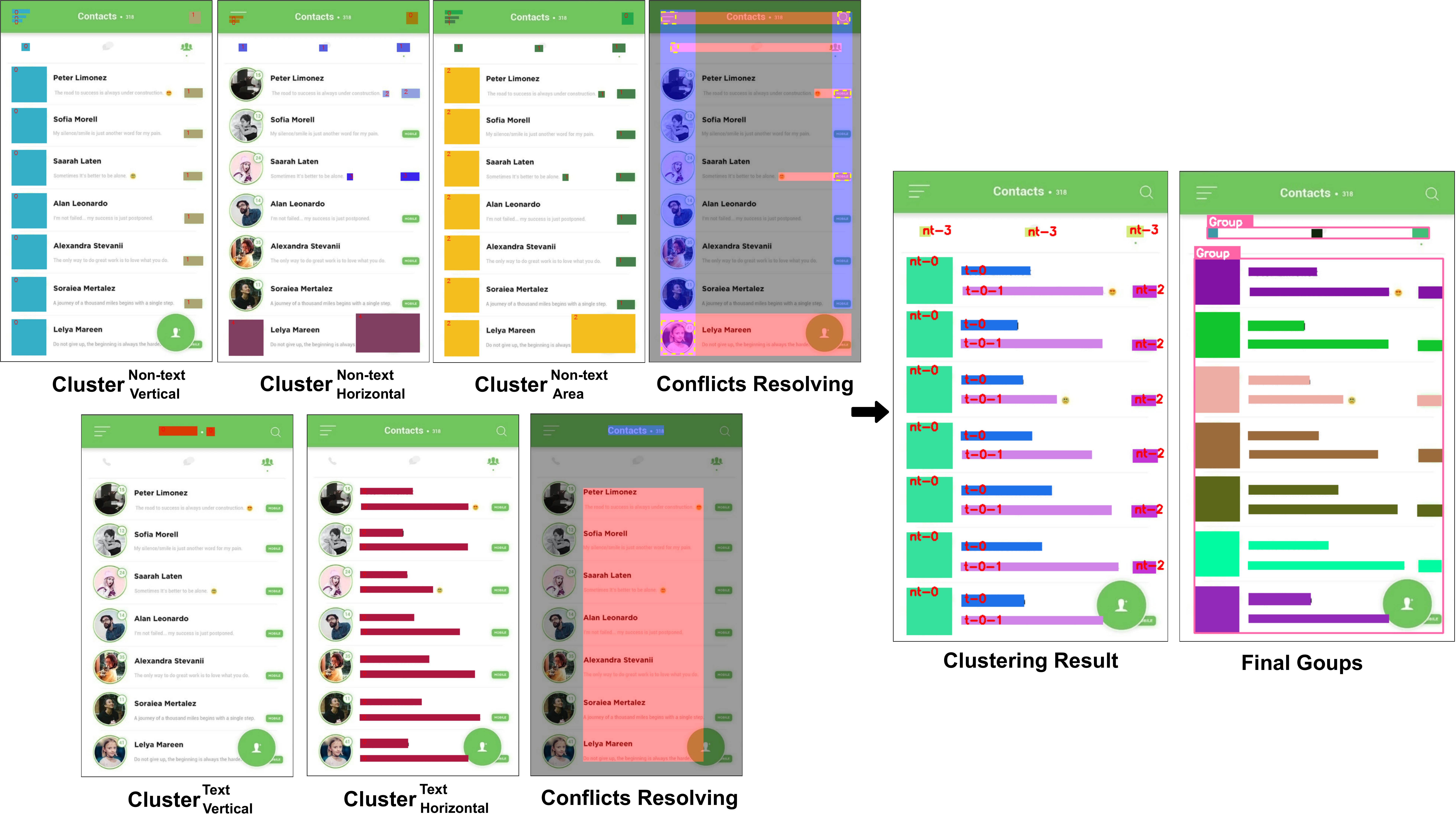}
    \caption{Widget clustering, cluster conflict resolving and final resulting groups in which we use the same color to paint the widgets in the same subgroup and highlight higher-order groups in pink boxes}
    \label{fig:groupingarch}
    \vspace{-4mm}
\end{figure}

Our approach uses the DBSCAN (Density-Based Spatial Clustering of Applications with Noise) algorithm~\cite{dbscan} to implement the clustering.
Intuitively, DBSCAN groups the points closely packed together (points with many nearby neighbors), while marking points whose distance from the nearest neighbor is greater than the maximum threshold as outliers. 
In the GUI widget clustering context, the point is the GUI widget, and the distance is the difference between the values of the widgets' attribute that the clustering is based on.
Figure~\ref{fig:groupingarch} illustrates the clustering process. 
For non-text widgets, our approach performs the clustering three times based on three attributes respectively.
It first clusters the widgets by $Center_x$ for the horizontal alignment, then by $Center_Y$ for the vertical alignment and finally by $area$.
These operations produce three clusters: $Cluster_{horizontal}^{non-text}$, $Cluster_{vertical}^{non-text}$ and $Cluster_{area}^{non-text}$.
Our approach then clusters the text widgets twice based on their top left corner point $(Top, Left)$ for left-justified (vertical) and top-justified (horizontal) alignment.
It produces the $Cluster_{horizontal}^{text}$ based on the texts' $Top$, and the $Cluster_{vertical}^{text}$ based on the texts' $Left$.
The resulting clusters are highlighted by different colors and numbers in Figure~\ref{fig:groupingarch}.
We only keep the clusters with at least two widgets and discard those with only one widget.


\subsubsection{Cluster Conflicts Resolving}
It is common that some widgets can be clustered into different clusters by different attributes, which causes cluster conflicts. 
For example, as illustrated in Figure~\ref{fig:groupingarch}, several non-text widgets (e.g.,  the bottom-left image) are both in a vertical cluster (marked in blue) and a horizontal cluster (marked in red).
The intersection of clusters illustrates the conflict.
The approach shall resolve such cluster conflicts to determine to which group the widget belongs.
This conflict-resolving step also complies with the similarity principle that suggests the widgets in the same perceptual group should share more similar properties.

The conflict resolving step first calculates the average widget areas of the groups to which the conflicting widget has been assigned.
In accordance with the similarity principle, the widget is more likely to be in a group whose average widget area is similar to the conflicting widget's area.
In addition, another observation is that repetitive widgets in a group have similar spacing between each other.
So for a widget that is clustered into multiple candidate groups, the approach checks the difference between the spacing of this widget and its neighboring widgets in a group and the average spacing between other widgets in that group.
It keeps the widget in the group where the conflicting widget has the largest widget-area similarity and the smallest spacing difference compared with other widgets in the group.
For example, the bottom-left image widget will be assigned to the vertical cluster rather than the horizontal one according to our conflict resolving criteria.
After conflict resolving, our approach produces the final widget clustering results as shown in the right part of Figure~\ref{fig:groupingarch}.
We use different colors and indices to illustrate the resulting non-text (nt) and text (t) clusters.
\begin{figure}
    \centering
    \includegraphics[width=0.47\textwidth]{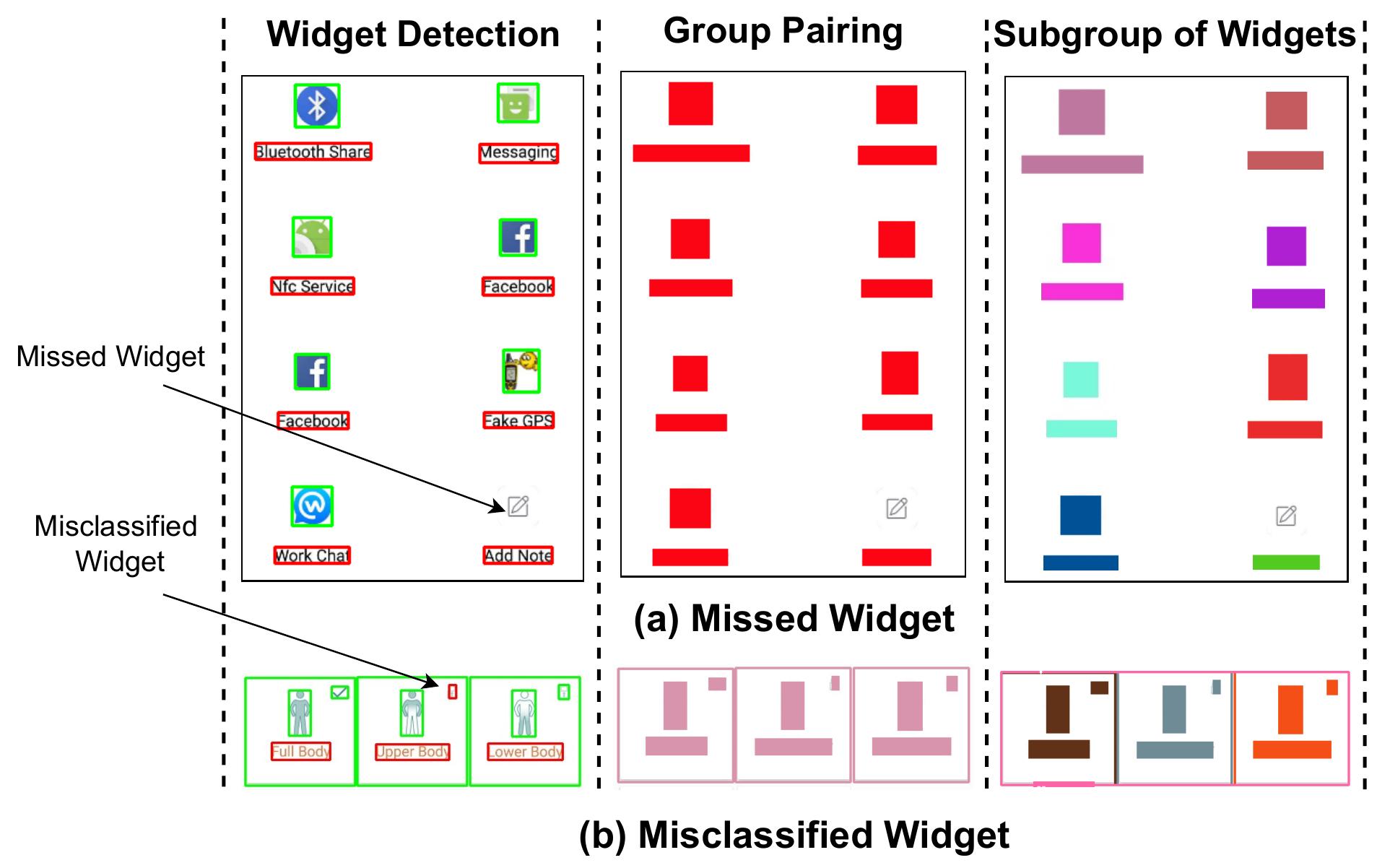}
    \caption{Examples of widget detection error correction. (1st column - green box: non-text; red-box: text; 2nd column - same color: higher-order perceptual group; 3rd column - same color: subgroup of widgets)}
    \label{fig:misdetection}
    \vspace{-4mm}
\end{figure}

\subsection{Proximity - Iterative Group Pairing}
So far, GUI widgets are aggregated into groups as per the connectedness and similarity principles.
Some groups are close to each other and similar in terms of the number and layout of the contained widgets, which may further form a larger perceptual group even though these groups may contain different types of widgets.
For example, in the clustering result of Figure~\ref{fig:groupingarch}, we can observe that the clusters \textit{nt-0}, \textit{t-0}, \textit{t-0-1} and \textit{nt-2} are proximate and have the same or similar number of widgets aligned in the same way.
We can see this feature intentionally or subconsciously and perceive them as in the same large group as a whole.
Gestalt psychology states that when people see an assortment of objects, they tend to perceive objects that are close to each other as a group \cite{wikipedia_gestalt}.
The close distance, also known as proximity, of elements is so powerful that it can override widget similarity and other factors that might differentiate a group of objects \cite{usertesting_7GP}.
Thus, the next step is based on widget clusters' proximity and composition similarity to pair the clusters into a larger group (i.e., layout block).

If two groups $Group_a$ and $Group_b$ are next (proximate) to each other (i.e., no other groups in between), and they contain the same number of widgets and the widgets in the $Group_a$ and the $Group_b$ share the same orientation (vertical or horizontal), our approach combines $Group_a$ and $Group_b$ into a larger block.
A widget in $Group_a$ and its closet widget in $Group_b$ will be paired and form a subgroup of widgets.
Our approach first combines the two proximate groups containing the same type of widgets, and then the groups containing different types of widgets.
The formed larger block can be iteratively combined with the proximate groups until no more proximate groups are available.

Sometimes there are different numbers of widgets in the two proximate groups but the two groups may still form one larger perceptual block.
For example, the cluster \textit{nt-2} in Figure~\ref{fig:groupingarch} has one less widget compared to \textit{nt-0}, \textit{t-0} and \textit{t-0-1} because the bottom widget in the right column is occluded by the float action button and thus missed by the detector.
Another common reason for the widget number difference is that widgets in a group may be set as invisible in some situations, and thus they do not appear visually.
Therefore, if the difference between the number of widgets in the two proximate groups is less than 4 (empirically determined from the ground-truth groups in our dataset), our approach also combines the two groups into a larger block.


As shown in the final groups in Figure~\ref{fig:groupingarch}, our approach identifies a set of perceptual groups (blocks), including the multitab at the top and the list in the main area.
Each list item is a combined widget of some non-text and text widgets (highlighted in the same color).
These perceptual groups encode the comprehension of the GUI structure into higher-order layout blocks that can be used in further processing and applications.

\subsection{Continuity - Detection Error Correction}

\begin{figure*}[t]
    \centering
    \includegraphics[width=0.98\textwidth]{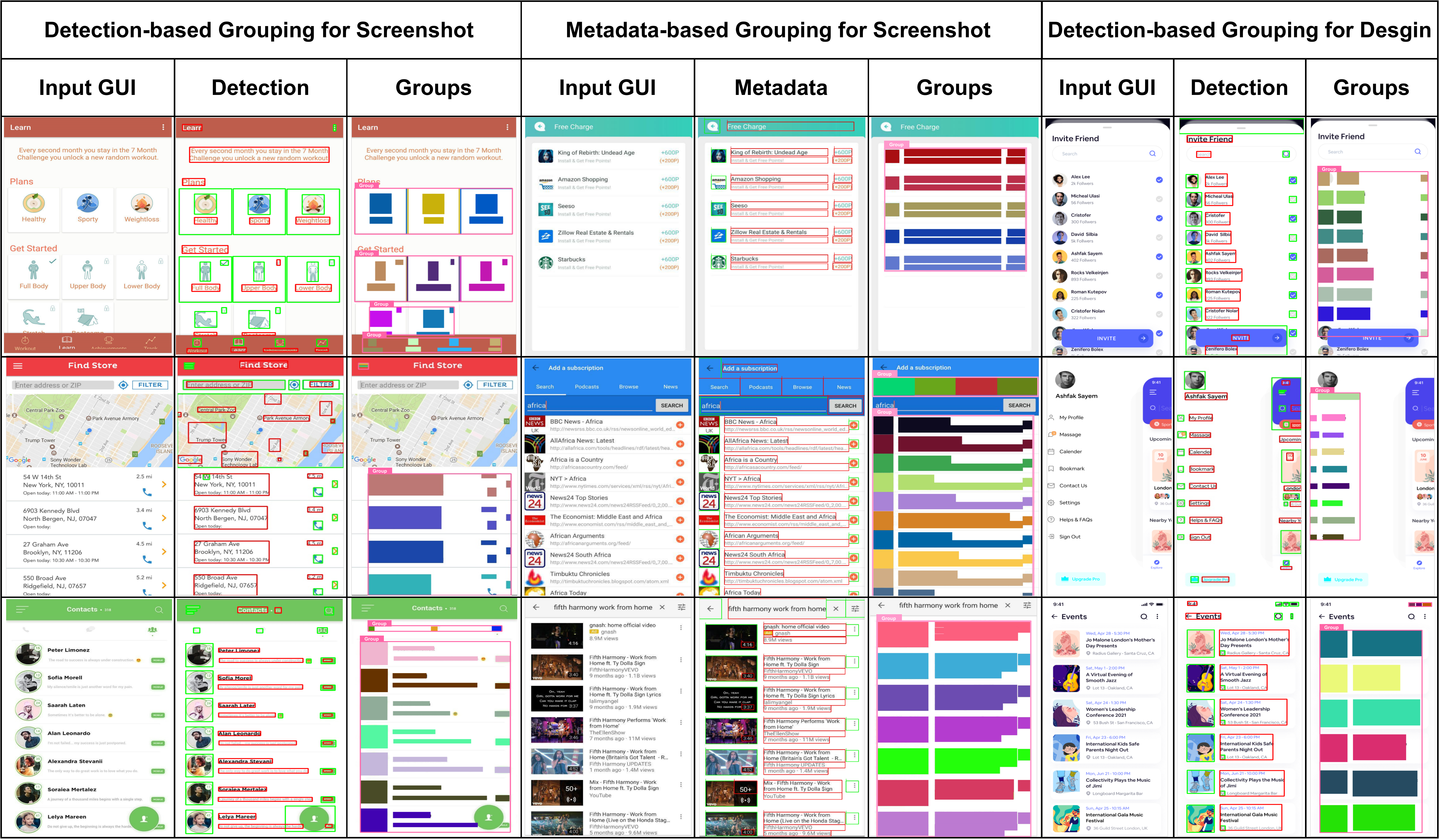}
    \caption{Examples of GUI widget detection and perceptual grouping results (red box - text widget, green box - non-text widget, pink box - perceptual group). Metadata-based means grouping the ground-truth widgets directly from GUI metadata.}
    \label{fig:results}
    \vspace{-2mm}
\end{figure*}

The GUI widget detector may make two types of detection errors - missed widgets and misclassified widgets.
Missed widgets means that the detector fails to detect some GUI elements on the GUI (e.g., the bottom-right icon in Figure~\ref{fig:misdetection}(a)).
Misclassified widgets refer to the widgets that the detector reports the wrong type, for example, the top-right small icon (i.e., a non-text widget) in the middle card in Figure~\ref{fig:misdetection}(b) is misclassified as a text widget due to an OCR error.

It is hard to recognize and correct these detection errors from the individual widget perspective, but applying the Gestalt continuity principle to expose such widget detection errors by contrasting widgets in the same perceptual groups can mitigate the issue.
The continuity principle states that elements arranged in a line or curve are perceived to be more related than elements not in a line or curve \cite{andyrutledge_3laws}.
Thus, some detection errors are likely to be spotted if a GUI area or a widget aligns with all the widgets in a perceptual group in a line but is not gathered into that group.

Our approach tries to identify and fix missed widgets as follows.
It first inspects the subgroups of widgets in a perceptual group and checks if the widgets in the subgroups are consistent in terms of the number and relative position of the contained widgets.
If a subgroup contains fewer widgets than its counterparts, then the approach locates the inconsistent regions by checking the relative positions and areas of other subgroups' widgets.
Next, the approach crops the located UI regions and uses the widget detector upon the cropped regions with relaxed parameters (i.e. double of the minimum area threshold for valid widgets) to try to identify the missed widget, if any.
For example, the tiny icon at the bottom right in Figure~\ref{fig:misdetection}(a) is missed because its area is so small that the detector regards it as a noisy region and hence discards it in the initial detection.
By analyzing the resulting perceptual group and its composition, our approach finds that seven of the eight subgroups have two widgets (marked in the same color), while the subgroup at the bottom right has only one widget.
It crops the area that may contain the missed widget according to the average sizes and average relative positions of the two widgets in the other seven subgroups.
The missed tiny icon can be recovered by detecting the widget with the relaxed valid-widget minimum area threshold in the missing area.

Our approach uses the exact mechanism that contrasts the subgroups to identify the misclassified widgets, but here it focuses on widget type consistency.
As shown in Figure~\ref{fig:misdetection}(b), our approach groups the three cards in a perceptual group.
By contrasting the widgets in the three cards, it detects that the middle card has text widgets at the top right corner, while the other two cards have a non-text widget at the same relative positions.
Based on the continuity principle, our approach re-classifies the top-right widget in the middle card as non-text with a majority-win strategy.

\section{Evaluation}
\label{sec:evaldetection}
We evaluate our approach in two steps: 
(1) examine the accuracy of our enhanced version of UIED and compare it with the original UIED~\cite{UIED-full-fse};
(2) examine the accuracy of our widget perceptual grouping approach and compare it with the state-of-the-art heuristic-based method Screen Recognition~\cite{screenrecognition}. 

\begin{table}[t]
	\centering
	\caption{Overall results of widget detection  (IoU $>$ 0.9)
	}
	\vspace{-1mm}
	\resizebox{0.48\textwidth}{10mm}{
    	\begin{tabular}{l | c c c | c c c}
    		\hline
    		& \multicolumn{3}{c|}{\textit{Our Enhanced Revision}} & \multicolumn{3}{c}{\textit{Original UIED}} \\
    		\hline
    		\textbf{Type} & \textbf{Precision} & \textbf{Recall} & \textbf{F1} & \textbf{Precision} & \textbf{Recall} & \textbf{F1} \\
    		\hline
    		Non-Text      & \textbf{0.589} & \textbf{0.823} & \textbf{0.687} & 0.431 & 0.469 & 0.449 \\
    		Text          & \textbf{0.678} & 0.693 & \textbf{0.686} & 0.402 & \textbf{0.720} & 0.516 \\
    		All Widgets          & \textbf{0.580} & \textbf{0.680} & \textbf{0.626} & 0.490 & 0.557 & 0.524 \\
    		\hline
    	\end{tabular}
	}
	\vspace{-3mm}
	\label{tab:detect}
\end{table}

\subsection{Accuracy of GUI Widget Detection}
\label{sec:detectioneval}

Compared with the original UIED~\cite{UIED-full-fse}, our GUI widget detector uses the latest Google OCR tool and improve the text and non-text widget merging by container analysis.
We evaluate GUI widget detection from the three perspectives: text widget detection, non-text widget detection and the final widget results after merging.
To be consistent with the evaluation setting in the UIED paper~\cite{UIED-full-fse}, we run experiments on the same Rico dataset of Android app GUIs~\cite{rico2} and regard the detected widgets whose intersection over union (IoU) with the ground truth widget is over 0.9 as true positive.
The ground-truth widgets are the leaf widgets (i.e., non-layout classes) extracted from the GUI's runtime view hierarchy.

Table~\ref{tab:detect} shows the widget detection performance of the enhanced and the original UIED.
Our enhanced version achieves a much higher recall (0.823) for non-text widgets than the original UIED (0.469), and meanwhile, it also improves the precision (0.589 over 0.431).
This significant improvement is due to the more intelligent container-aware merging of text and non-text widgets by our enhanced version.
As the original UIED is container-agnostic, it erroneously discards many valid widgets contained in other widgets as noise.
For GUI text, the Google OCR tool used in the enhanced version achieves much higher precision (0.678) than the EAST model used in the original UIED (0.402), with a slight decrease in recall (0.693 versus 0.720).
The improvements in both text and non-text widgets result in a much better overall performance (0.626 F1 by the enhanced version versus 0.524 by the original UIED).

\subsection{Perceptual Grouping Performance}
\label{sec:groupingperformance}
We evaluate our perceptual grouping approach on both Android app GUIs and UI design prototypes.
Figure ~\ref{fig:results} shows some perceptual grouping results by our approach.
These results show our approach can reliably detect GUI widgets and infer perceptual groups for diverse visual and layout designs.

\subsubsection{Datasets}
\label{dataset}
\begin{figure}
    \centering
    \includegraphics[width=0.47\textwidth]{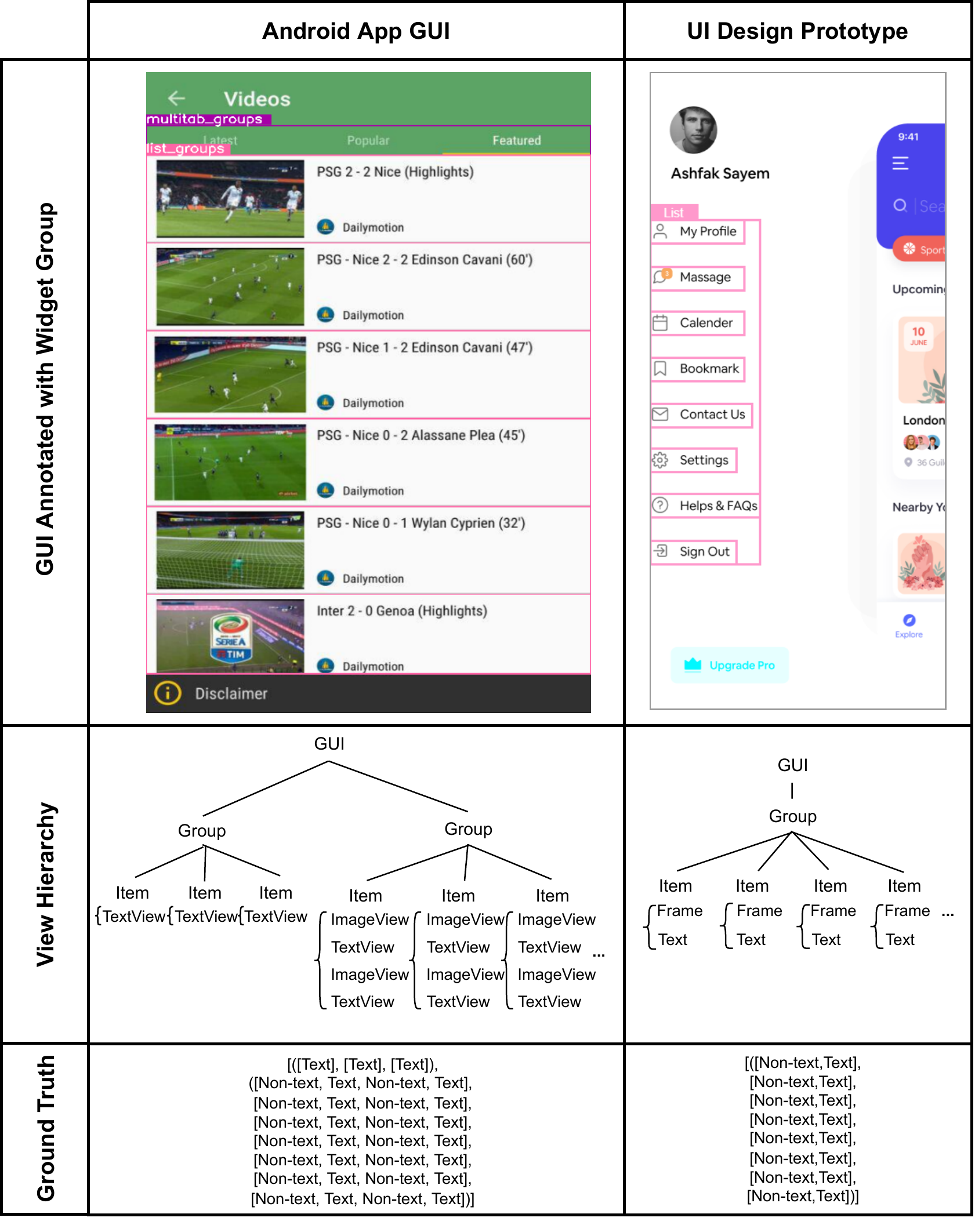}
    \caption{Examples of Android app GUI and UI design prototype, view hierarchy and ground truth}
    \label{fig:annotation}
    \vspace{-4mm}
\end{figure}

Our approach simulates how humans perceive the GUI structure and segment a GUI into blocks of widgets according to the Gestalt principles of grouping.
To validate the recognized blocks, we build the ground-truth dataset from two sources: Android apps and UI design prototypes.
The ground truth annotates the widget groups according to the GUI layout and the widget styles and properties as shown in Figure~\ref{fig:annotation}.

\textbf{Android App GUI Dataset}
\label{sec:datasetrico}
The ground truth of widget groups can be obtained by examining the layout classes used to group other widgets in the implementations.
However, as shown in Figure~\ref{fig:rico-wrong}, the layout classes do not always correspond to the perceptual groups of GUI widgets.
Therefore, we cannot use the GUI layout classes directly as the ground truth.
Instead, we first search the GUIs in the Rico dataset of Android app GUIs~\cite{rico2} that use certain Android layout classes that may contain a group of widgets (e.g., ListView, FrameLayout, Card, TabLayout).
Then we manually examine the candidate GUIs to filter out those whose use of layout classes has obvious violations against the Gestalt principles.
Furthermore, the Rico dataset contains many highly-similar GUI screenshots for an application.
To increase the visual and layout diversity of GUIs in our dataset, we limit up to three distinct GUI screenshots per application.
Distinction is determined by the number and type of GUI widgets and the GUI structure.
We obtain 1091 GUI screenshots from 772 Android applications.
Using this dataset, we evaluate both detection-based and metadata-based grouping.
Detection-based grouping processes the detected widgets, while metadata-based grouping uses the widgets obtained from the GUI metadata (i.e., assumes the perfect widget detection).


\textbf{UI Design Prototypes}
\label{sec:datasetfigma}
We collect 20 UI design prototypes shared on a popular UI design website (Figma~\cite{figma}).
These UI design prototypes are created by professional designers for various kinds of apps and receive more than 200 likes.
This small set of UI design prototypes demonstrates how professional designers structure the GUIs and group GUI widgets from the design rather than the implementation perspective.
As a domain-independent tool, Figma supports only elementary visual elements (i.e., text, image and shape).
Designers can create any widgets using these elementary visual elements.
Due to the lack of explicit and uniform widget metadata in the Figma designs, we evaluate only the detection-based grouping on these UI design prototypes.

\subsubsection{Metrics}

The left part of Figure~\ref{fig:annotation} shows an example in our Android app GUI dataset.
We see that the layout classes (e.g., ListView, TabLayout) in the view hierarchy map to the corresponding perceptual groups.
In our dataset, specific layout classes are generalized to blocks, as we only care about generic perceptual groups in this work.
Following the work~\cite{chunyangui2code} for generating GUI component hierarchy from UI image, we adopt the sequence representation of a GUI component hierarchy.
Through depth-first traversal, a view hierarchy can be transformed into a string of GUI widget names and brackets (``[]'' and ``()'' corresponding to the blocks).
This string represents the ground-truth perceptual groups of an app GUI.
As discussed in Section~\ref{sec:widgetdetection}, perceptual grouping is based on the widgets' positional and visual properties while the actual classes of non-text widgets are not necessary.
Thus, the ground-truth string only has two widget types: Text and Non-text.
Specifically, it converts TextView in the view hierarchy to text and all other classes as Non-text.
Similarly, as shown in the right part of Figure~\ref{fig:annotation}, the designers organize texts and non-text widgets (images or shape compositions referred to as frames) into a hierarchy of groups. 
Based on the design's group hierarchy, we output the ground-truth string of perceptual groups.
The perceptual groups ``perceived'' by our approach are output in the same format for comparison.

We compute the Levenshtein edit distance between the two strings of a ground-truth block and a perceived block.
The Levenshtein edits inform us of the specific mismatches between the two blocks, which is important to understand and analyze grouping mistakes.
If the edit distance between the ground-truth block and the perceived block is less than a threshold, we regard the two blocks as a candidate match.
We determine the optimal matching between the string of ground-truth blocks and the string of the perceived blocks by minimizing the overall edit distance among all candidate matches.
If a perceived group matches a ground-truth group, it is a true positive (TP), otherwise a false positive (FP).
If a ground-truth group does not match any perceived group, it is a false negative (FN).
Based on the matching results, we compute: (1) precision (\textit{TP/(TP+FP)}) (2) recall (\textit{TP/(TP+FN)}), and (3) F1-score (\textit{(2*precision*recall) / (precision+recall)})

\subsubsection{Performance on Android App GUIs}

\begin{figure}[t]
    \centering
    \includegraphics[width=0.47\textwidth]{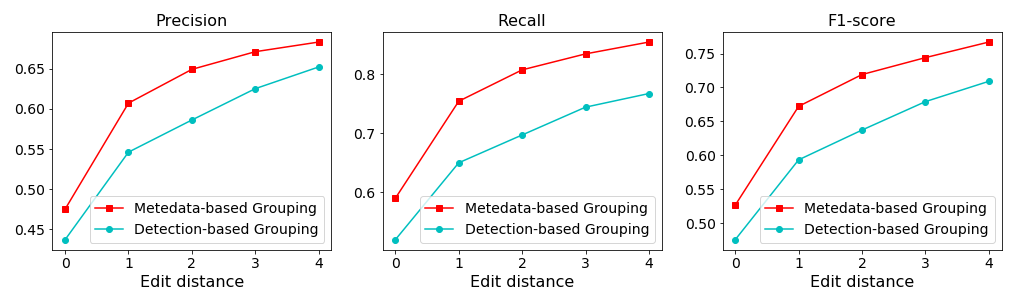}
    \caption{Performance at different edit distance thresholds}
    \label{fig:groupperformance}
    \vspace{-4mm}
\end{figure}



We experiment with five edit distance thresholds (0-4).
The distance 0 means the two blocks have the perfect match, and the distance 4 means as long as the unmatched widgets in the two blocks are no more than 4, the two blocks can be regarded as a candidate match.
As shown in Figure~\ref{fig:groupperformance}, for detection-based grouping, the precision, recall and F1-score is 0.437, 0.520 and 0.475 at the distance 0.
As the distance threshold increases (i.e., the matching criterion relaxes), the precision, recall and F1-score keeps increasing to 0.652, 0.776 and 0.709 at the distance threshold 4.
As shown in Figure~\ref{fig:groupperformance} and Table~\ref{tab:groupperformance}, the metadata-based grouping with the ground-truth GUI widgets achieves a noticeable improvement over the detection-based grouping with the detected widgets, in terms of all three metrics, especially for recall.
This suggests that improving GUI widget detection will positively improve the subsequent perceptual grouping.

As the examples in Figure~\ref{fig:results} show, our approach can not only accurately process GUIs with clear structures (e.g., the first row), but it can also process GUIs with large numbers of widgets that are placed in a packed way (e.g., the second and third rows).
Furthermore, our approach is fault-tolerant to GUI widget detection errors to a certain extent, for example, the second row of detection-based grouping for screenshot and design.
The map and the pushed-aside partial GUI result in many inaccurately detected GUI widgets in these two cases.
However, our approach still robustly recognizes the proper perceptual groups.



We compare our approach with the heuristic-based grouping method (Screen Recognition) proposed in Zhang et.al.~\cite{screenrecognition} (which received the distinguished paper award at CHI2021).
The results in Table~\ref{tab:groupperformance} shows that Screen Recognition can hardly handle visually and structurally complicated GUIs based on a few ad-hoc and rigid heuristics.
Its F1 score is only 0.092 on the detected widgets and 0.123 on the ground-truth widgets.
This is because its heuristics are designed for only some fixed grouping styles such as cards and multi-tabs.
In contrast, our approach is designed to fulfil generic Gestalt principles of grouping.

\begin{table}[t]
    \centering
    \caption{Performance comparison (edit distance$\leq$1)}
    \resizebox{0.47\textwidth}{10mm}{
        \begin{tabular}{l l c c c c}
             \textbf{Widgets} & \textbf{Approach} & \textbf{\#Bock} & \textbf{Precision} & \textbf{Recall} & \textbf{F1}\\
             \hline
             \multirow{2}{*}{Metadata} & Our Approach & 1,465 & \textbf{0.607} & \textbf{0.754} & \textbf{0.672}  \\
             & Screen Recog & 1,038 & 0.131 & 0.116 & 0.123 \\
             \hline
             \multirow{2}{*}{Detection} & Our Approach & 1,260 & 0.546 & 0.650 & 0.593 \\
             & Screen Recog & 992 & 0.103 & 0.083 & 0.092 \\ 
             \hline
        \end{tabular}
    }
    \label{tab:groupperformance}
    \vspace{-4mm}
\end{table}

\begin{figure}[t]
    \centering
    \includegraphics[width=0.47\textwidth]{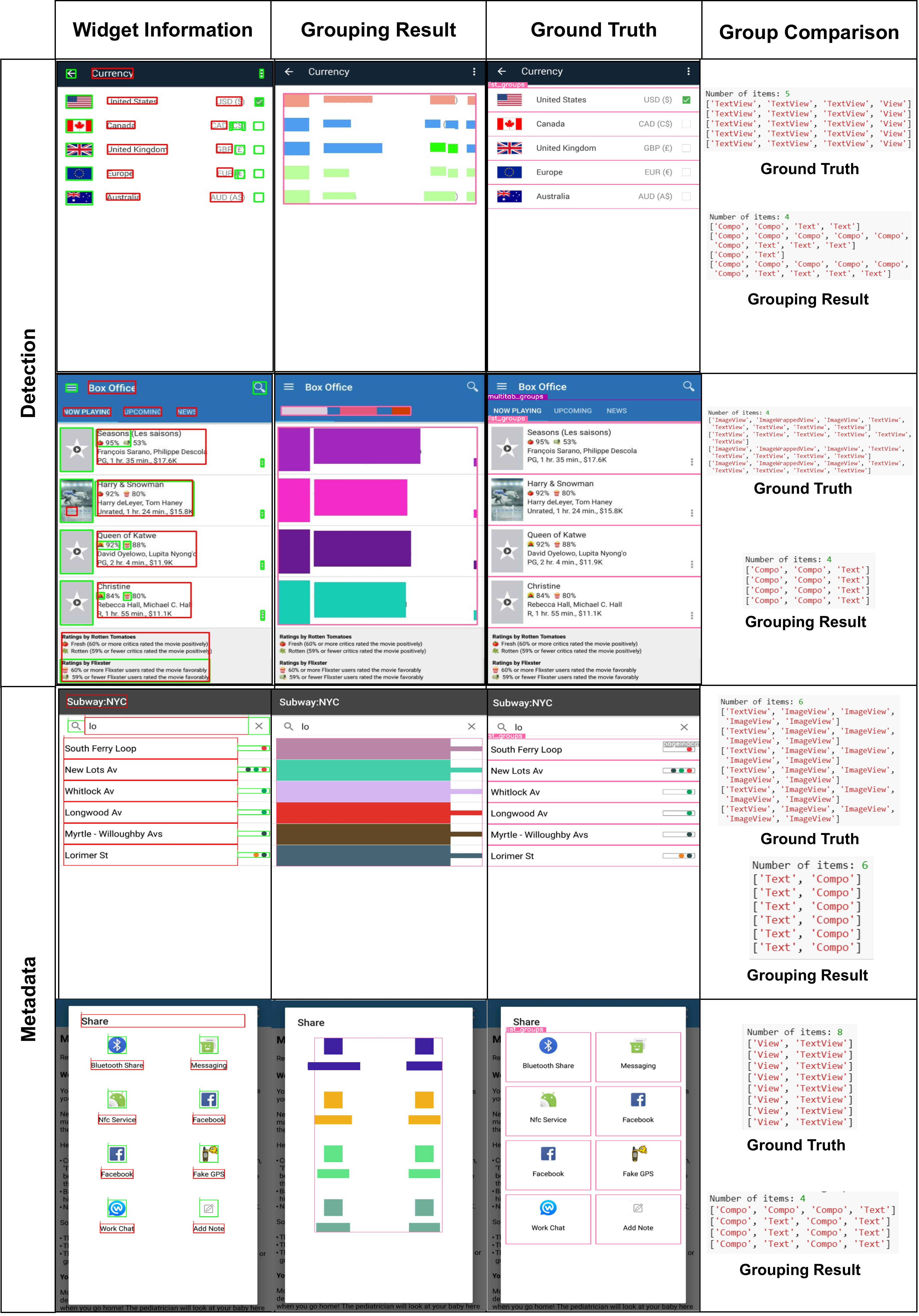}
    \caption{Typical causes of grouping mistakes (red box - text widget, green box - non-text widget, pink box - perceptual group, red dashed box - unmatched ground-truth widget)}
    \label{fig:errors}
    \vspace{-4mm}
\end{figure}

We manually inspect the grouping results by our approach against the ground-truth groups to identify the potential improvements.
Figure~\ref{fig:errors} presents four typical cases that cause the perceived groups to be regarded as incorrect.
For the detection-based grouping, the major issue is GUI widget over-segmentation (a widget is detected as several widgets) or under-segmentation (several widgets are detected as one widget).
In the first row, the detector segments the texts on the right side of the GUI into several text and non-text widgets.
As indicated by the same color in the Grouping Result column, our approach still successfully partitions the widgets on the same row into a block, and recognizes the large group containing these row blocks.
But as shown in the Group Comparison column, one widget in the second, third and fourth detected blocks do not match those in the corresponding ground-truth blocks.
In the second row, the GUI widget detector merges close-by multi-line texts as a single text widget, while these text widgets are separate widgets in the ground truth.
Again, our approach recognizes the overall perceptual groups correctly, but the widgets in the corresponding blocks do not completely match.

While using the ground-truth widgets from the GUI metadata to mitigate the GUI widget misdetection, the grouping results see the improvement but suffer from two other problems.
First, the widgets in the metadata contain some widgets that are visually occluded or hidden.
The third row in Figure~\ref{fig:errors} illustrates this problem, where some widgets are actually occluded behind the menu on the left, but they are still available in the runtime metadata and are extracted as the ground-truth widgets.
This results in a completely incorrect grouping.
The issue of widget occlusion or modal window could be mitigated as follows: 
train an image classifier to predict the presence of widget occlusion or modal window, then follow the figure-ground principle~\cite{gelstal-foreground} to separate foreground modal window from the background, and finally detect the perceptual groups on the separated model window.
Second, alternative ways exist to partition the widgets into groups.
For example, for the GUI in the fourth row, the ground truth contains eight blocks, each of which has one image and one text while our grouping approach partitions these blocks into four rows of a large group, and each row contains two blocks (as indicated by the same color in Grouping Result).
Perceptually, both ways are acceptable but the group differences cause the grouping result by our approach to be regarded as incorrect.


\subsubsection{Performance on UI Design Prototypes}

Tested on the 20 UI design prototypes, our approach achieves the precision of 0.750, the recall 0.818 and the F1-score 0.783.
The third column in Figure~\ref{fig:results} shows some results of our grouping approach for the UI design prototypes, where we see it is able to infer the widget groups well for different styles of GUI designs.
GUI widget detection is more robust on UI design prototypes due to the more accurate GUI widget detection, which leads to the improvement of the subsequent grouping of detected widgets.
As shown in Figure~\ref{fig:results}, the widgets in a UI design prototype is usually scattered, while the real app GUIs are packed.
Both GUI widget detection and perceptual grouping become relatively easier on less packed GUIs.

\subsubsection{Processing Time}
As the GUI widget grouping can be used as a part of various automation tasks such as automated testing, the runtime performance can be a concern.
We record the processing time while running our approach over the dataset to get a sense of its efficiency.
Our experiments run on a machine with Windows 10 OS, Intel i7-7700HQ CPU, and 8GB memory.
Our approach comprises two major steps: widget detection and perceptual grouping. 
We improved and refactored the original UIED to boost the run performance of the widget detection, and now it takes an average of ~1.1s to detect the widgets in a GUI, which significantly exceeds the original UIED that takes on average ~9s per GUI.
The grouping process is also efficient, which takes an average of 0.6s to process a GUI.
In total, the average processing time of the entire approach is ~1.7s per GUI image.
Furthermore, as our approach does not involve any deep learning techniques, it does not require advanced computing support such as GPU.
\section{Related Work}



Our work falls into the area of reverse-engineering the hidden attributes of GUIs from pixels.
There are two lines of closely related work: GUI widget detection and GUI-to-text/code generation.

GUI widget detection is a special case of object detection~\cite{selectivesearch, frcnn, yolo}.
Earlier work~\cite{remaui} uses classic computer vision (CV) algorithms (e.g., Canny edge and contour analysis) to detect GUI widgets.
Recently, White et al.~\cite{whiteyolo} apply a popular object detection model YOLOv2~\cite{yolo} to detect GUI widgets in GUI images for random GUI testing.
Feng et al.~\cite{galleryDC} apply Faster RCNN~\cite{frcnn} to obtain GUI widgets from app screenshots and construct a searchable GUI widget gallery.
Chen et al.~\cite{chen2022towards} proposed an approach to complete icon labeling in mobile applications.
A recent study by Xie et al.~\cite{UIED-full-fse} shows that both classic CV algorithms and recent deep learning models have limitations when applied to GUI widget detection, which has different visual characteristics and detection goals from natural scene object detection.
They design a hybrid method UIED inspired by the figure-ground~\cite{gelstal-foreground} characteristic of GUI, which achieves the start-of-the-art performance for GUI widget detection.

GUI-to-text/code generation also receives much attention.
To improve GUI accessibility, Chen et al.~\cite{chenjieshan2020ase} propose a transformer-based image captioning model for producing labels for icons.
To implement GUI view hierarchy, REMAUI~\cite{remaui} infers three Android-specific layouts (LinearLayout, FrameLayout and ListView) based on hand-craft rules to group widgets.
Recently, Screen Recognition~\cite{screenrecognition} develops some heuristics for inferring tabs and bars.
However, these heuristic-based widget grouping methods cannot handle visually and structurally complicated GUI designs (e.g., nested perceptual groups like a grid of cards).
Alternatively, image captioning models~\cite{imgcaption1,deeplearning_Nature} have been used to generate GUI view hierarchy from GUI images~\cite{beltramelli2018pix2code, chunyangui2code}.
Although these image-captioning based methods get rid of hard-coded heuristics, they suffer from GUI data availability and quality issues (as discussed in Introduction and illustrated in Figure~\ref{fig:rico-wrong}).
These methods also suffer from code redundancy and no explicit image-code traceability issues (see Section~\ref{sec:modulargui2code}).
The perceptual groups recognized by our approach could help to address these issues.

None of the existing GUI widget detection and GUI-to-code approaches solve the perceptual grouping problem in a systematic way as our approach does.
ReDraw~\cite{redraw} and FaceOff~\cite{faceoff} solves the layout problem by finding in the codebase the layouts containing similar GUI widgets.
Some other methods rely on source code or specific layout algorithm (e.g., Android RelativeLayout) to synthesize modular GUI code or layout~\cite{inferui, reusbale_code} or infer GUI duplication~\cite{nearduplicate1}.
All these methods are GUI implementation-oriented, and hard to generalize for other application scenarios such as UI design search, UI automation, robotic GUI testing or accessibility enhancement.
In contrast, our approach is based on domain-independent Gestalt principles and is application-independent, so it can support different downstream SE tasks (see Section~\ref{sec:applications}).

In the computer vision community, some machine learning techniques  ~\cite{zellers2018neural, knyazev2020graph, yang2018graph} have been proposed to predict structure in the visual scene, i.e., so-called scene graphs.
These techniques can infer the relationships between objects detected in an image and describe these relationships by triplets (<subject, relation, object>).
However, such relationship triplets cannot represent complex GUI widget relations in perceptual groups.
Furthermore, these techniques also require sufficient high-quality data for model training, which is a challenging issue for GUIs.

\section{Perceptual Grouping Applications}
\label{sec:applications}

Our perceptual grouping method fills in an important gap for automatic UI understanding.
Perceptual groups, together with elementary widget information, would open the door to some innovative applications in software engineering domain.


\subsection{UI Design Search}

UI design is a highly creative activity.
The proliferation of UI design data on the Internet enables data-driven methods to learn UI designs and obtain design inspirations~\cite{rico, rico2,chen2020lost}.
However, this demands effective UI design search engines.
Existing methods often rely on the GUI metadata, which limits their applicability as most GUI designs exist in only pixel format.
GalleryDC~\cite{galleryDC} builds a gallery of GUI widgets and infer elementary widget information (e.g., size, primary color) to help widget search.
Unfortunately, this solution does not apply to the whole and complex UIs.
Chen et al.~\cite{Chen_uisearch} and Rico~\cite{rico2} use image auto-encoder to extract image features through self-supervised learning, which can be used to find visually similar GUI images.
However, the image auto-encoder encodes only pixel-level features, but is unaware of GUI structure, which is very critical to model and understand GUIs.
As such, given the GUI in the left of Figure~\ref{fig:rico-wrong}, these auto-encoder based methods may return a GUI like the one on the right of Figure~\ref{fig:rico-wrong}, because both GUIs have rich graphic features and some textural features.
Unfortunately, such search results are meaningless, because they bear no similarity in terms of GUI structure and perceptual groups of GUI widgets.
Our approach can accurately infer perceptual groups of GUI widgets from pixels.
Based on its perceptual grouping results, a UI design search would become structure-aware, and finds not only visually but also structurally similar GUIs.
For example, a structure-aware UI design search would return a GUI like the one in 2nd-row-1st-column of Figure~\ref{fig:results} for the left GUI in Figure~\ref{fig:rico-wrong}.

\subsection{Modular GUI-to-Code Generation}
\label{sec:modulargui2code}
Existing methods for GUI-to-code generation either use hand-craft rules or specific layout algorithms to infer some specific implementation layout~\cite{remaui, inferui}, or assume the availability of a codebase to search layout implementations~\cite{redraw, faceoff}.
Image-captioning based GUI-to-Code methods~\cite{beltramelli2018pix2code, chunyangui2code,feng2021auto,feng2022auto} are more flexible as they learn how to generate GUI view hierarchy from GUI metadata (if available).
However, the nature of image captioning is to just describe the image content, but it is completely unaware of GUI structure during the code generation.
As such, the generated GUI code is highly redundant for repetitive GUI blocks.
For example, for the card-based GUI design in Figure~\ref{fig:examples}(a), it will generate eight pieces of repetitive code, one for each of the eight cards.
This type of generated code is nothing like the modular GUI code developers write.
So it has little practicality.
Another significant limitation of image captioning is that the generated GUI layouts and widgets have no connection to the corresponding parts in the GUI image.
For a GUI with many widgets (e.g., those in the 2nd and 3rd rows in Figure~\ref{fig:results}), it would be hard to understand how the generated code implements the GUI.
With the support of our perceptual grouping, GUI-to-code generation can encapsulate the widget grouping information into the code generation process and produce much less redundant and more modular, reusable GUI code (e.g., extensible card component).

\subsection{UI Automation}

Automating UI understanding from pixels can support many UI automation tasks.
A particular application of UI automation in software engineering is automatic GUI testing.
Most existing methods for automatic GUI testing rely on OS or debugging infrastructure~\cite{guo2019sara, sahin2019randr, androidmonkey, UIautomator}.
In recent years, computer vision methods have also been used to support non-intrusive GUI testing~\cite{whiteyolo, roscript,feng2022gifdroid1,feng2022gifdroid2}.
However, these methods only work at the GUI widget level through either traditional widget detection~\cite{remaui} or deep learning models like Yolo~\cite{yolo}.
Furthermore, they only support random testing, i.e., random interactions with some widgets.
Some studies~\cite{issta2019learninguielementinteractions, deepgui, chen2022extracting} show that GUI testing would be more effective if the testing methods were aware of more likely interactions.
They propose deep learning methods to predict such likely interactions.
However, the learning is a completely black box.
That is, they can predict where on the GUI some actions could be applied, but they do not know what will be operated and why so.
Our approach can inform the learning with higher-order perceptual groups of GUI widgets so that the model could make an explainable prediction, for example, scrolling is appropriate because this part of GUI displays a list of repetitive blocks.
It may also guide the testing methods to interact with the blocks in a perceptual group in an orderly manner, and ensure all blocks are tested without unnecessary repetitions.
Such support for UI automation would also enhance the effectiveness of screen readers which currently heavily rely on accessibility metadata and use mostly elementary widget information.

\section{Conclusion and Future Work}

This paper presents a novel approach for recognizing perceptual groups of GUI widgets in GUI images.
The approach is designed around the four psychological principles of grouping - connectedness, similarity, proximity and continuity.
To the best of our knowledge, this is the first unsupervised, automatic UI understanding approach with a systematic theoretical foundation, rather than relying on ad-hoc heuristics or model training with GUI metadata.
Through the evaluation of both mobile app GUIs and UI design prototypes, we confirm the high accuracy of our perceptual grouping method for visually and structurally diverse GUIs.
Our approach fills the gap of visual intelligence between the current widget-level detection and the whole-UI level GUI-to-code generation.
As a pixel-only and application-independent approach, we envision our approach could enhance many downstream software engineering tasks with the visual understanding of GUI structure and perceptual groups, such as structure-aware UI design search, modular and reusable GUI-to-code generation, and layout-sensitive UI automation for GUI testing and screen reader.
Although our current approach achieves very promising performance, it can be further improved by dealing with widget occlusion or modal window.
Moreover, we will investigate semantic grouping that aims to recognize both interaction and content semantics of perceptual groups. 

\section*{Acknowledgements}
This research was partially funded by Data61-ANU Collaborative
Research Project.
Chunyang is partially supported by Facebook/Meta Gift grant.

\bibliographystyle{ACM-Reference-Format}
\bibliography{reference}


\begin{thebibliography}{60}


\ifx \showCODEN    \undefined \def \showCODEN     #1{\unskip}     \fi
\ifx \showDOI      \undefined \def \showDOI       #1{#1}\fi
\ifx \showISBNx    \undefined \def \showISBNx     #1{\unskip}     \fi
\ifx \showISBNxiii \undefined \def \showISBNxiii  #1{\unskip}     \fi
\ifx \showISSN     \undefined \def \showISSN      #1{\unskip}     \fi
\ifx \showLCCN     \undefined \def \showLCCN      #1{\unskip}     \fi
\ifx \shownote     \undefined \def \shownote      #1{#1}          \fi
\ifx \showarticletitle \undefined \def \showarticletitle #1{#1}   \fi
\ifx \showURL      \undefined \def \showURL       {\relax}        \fi
\providecommand\bibfield[2]{#2}
\providecommand\bibinfo[2]{#2}
\providecommand\natexlab[1]{#1}
\providecommand\showeprint[2][]{arXiv:#2}

\bibitem[\protect\citeauthoryear{??}{gel}{[n.d.]}]%
        {gelstal-foreground}
 \bibinfo{year}{[n.d.]}\natexlab{}.
\newblock \bibinfo{title}{7 Gestalt Principles of Visual Perception: Cognitive
  Psychology for UX: UserTesting Blog}.
\newblock
\newblock
\urldef\tempurl%
\url{https://www.usertesting.com/blog/gestalt-principles#figure}
\showURL{%
\tempurl}


\bibitem[\protect\citeauthoryear{??}{wix}{[n.d.]}]%
        {wix}
 \bibinfo{year}{[n.d.]}\natexlab{}.
\newblock \bibinfo{title}{Free website builder: Create a free website}.
\newblock
\newblock
\urldef\tempurl%
\url{http://www.wix.com/}
\showURL{%
\tempurl}


\bibitem[\protect\citeauthoryear{??}{goo}{[n.d.]a}]%
        {googleaccess}
 \bibinfo{year}{[n.d.]}\natexlab{a}.
\newblock \bibinfo{title}{Get started on Android with TalkBack - Android
  Accessibility Help}.
\newblock
\newblock
\urldef\tempurl%
\url{https://support.google.com/accessibility/android/answer/6283677?hl=en}
\showURL{%
\tempurl}


\bibitem[\protect\citeauthoryear{??}{fig}{[n.d.]}]%
        {figma}
 \bibinfo{year}{[n.d.]}\natexlab{}.
\newblock \bibinfo{title}{the collaborative interface design tool.}
\newblock
\newblock
\urldef\tempurl%
\url{https://www.figma.com/}
\showURL{%
\tempurl}


\bibitem[\protect\citeauthoryear{??}{and}{[n.d.]}]%
        {androidmonkey}
 \bibinfo{year}{[n.d.]}\natexlab{}.
\newblock \bibinfo{title}{UI/Application Exerciser Monkey : Android
  Developers}.
\newblock
\newblock
\urldef\tempurl%
\url{https://developer.android.com/studio/test/monkey#:~:text=The Monkey is a
  program,a random yet repeatable manner.}
\showURL{%
\tempurl}


\bibitem[\protect\citeauthoryear{??}{goo}{[n.d.]b}]%
        {googleocr}
 \bibinfo{year}{[n.d.]}\natexlab{b}.
\newblock \bibinfo{title}{Vision AI | Derive Image Insights via ML | Cloud
  Vision API}.
\newblock
\newblock
\urldef\tempurl%
\url{https://cloud.google.com/vision/}
\showURL{%
\tempurl}


\bibitem[\protect\citeauthoryear{??}{ges}{2006}]%
        {gestaltpsychology}
 \bibinfo{year}{2006}\natexlab{}.
\newblock \bibinfo{booktitle}{\emph{"Gestalt psychology". Britannica concise
  encyclopedia}}.
\newblock \bibinfo{publisher}{Britannica Digital Learning}.
\newblock
\showISBNx{9781593394929}


\bibitem[\protect\citeauthoryear{??}{app}{2021}]%
        {appleaccess}
 \bibinfo{year}{2021}\natexlab{}.
\newblock \bibinfo{title}{Accessibility - Vision}.
\newblock
\newblock
\urldef\tempurl%
\url{https://www.apple.com/accessibility/vision/}
\showURL{%
\tempurl}


\bibitem[\protect\citeauthoryear{??}{wik}{2021}]%
        {wikipedia_gestalt}
 \bibinfo{year}{2021}\natexlab{}.
\newblock \bibinfo{title}{Gestalt psychology}.
\newblock
\newblock
\urldef\tempurl%
\url{https://en.wikipedia.org/wiki/Gestalt_psychology#cite_note-1}
\showURL{%
\tempurl}


\bibitem[\protect\citeauthoryear{??}{UIa}{2021}]%
        {UIautomator}
 \bibinfo{year}{UIAutomator, 2021}\natexlab{}.
\newblock
  \bibinfo{howpublished}{\url{https://developer.android.com/training/testing/ui-automator}}.
\newblock


\bibitem[\protect\citeauthoryear{Bajammal, Mazinanian, and Mesbah}{Bajammal
  et~al\mbox{.}}{2018}]%
        {reusbale_code}
\bibfield{author}{\bibinfo{person}{Mohammad Bajammal}, \bibinfo{person}{Davood
  Mazinanian}, {and} \bibinfo{person}{Ali Mesbah}.}
  \bibinfo{year}{2018}\natexlab{}.
\newblock \showarticletitle{Generating Reusable Web Components from Mockups}.
  In \bibinfo{booktitle}{\emph{Proceedings of the 33rd ACM/IEEE International
  Conference on Automated Software Engineering}} (Montpellier, France)
  \emph{(\bibinfo{series}{ASE 2018})}. \bibinfo{address}{New York, NY, USA},
  \bibinfo{pages}{601–611}.
\newblock
\showISBNx{9781450359375}
\urldef\tempurl%
\url{https://doi.org/10.1145/3238147.3238194}
\showDOI{\tempurl}


\bibitem[\protect\citeauthoryear{Beltramelli}{Beltramelli}{2018}]%
        {beltramelli2018pix2code}
\bibfield{author}{\bibinfo{person}{Tony Beltramelli}.}
  \bibinfo{year}{2018}\natexlab{}.
\newblock \showarticletitle{pix2code: Generating code from a graphical user
  interface screenshot}. In \bibinfo{booktitle}{\emph{Proceedings of the ACM
  SIGCHI Symposium on Engineering Interactive Computing Systems}}.
  \bibinfo{pages}{1--6}.
\newblock


\bibitem[\protect\citeauthoryear{Bielik, Fischer, and Vechev}{Bielik
  et~al\mbox{.}}{2018}]%
        {inferui}
\bibfield{author}{\bibinfo{person}{Pavol Bielik}, \bibinfo{person}{Marc
  Fischer}, {and} \bibinfo{person}{Martin Vechev}.}
  \bibinfo{year}{2018}\natexlab{}.
\newblock \showarticletitle{Robust Relational Layout Synthesis from Examples
  for Android}.
\newblock \bibinfo{journal}{\emph{Proc. ACM Program. Lang.}}
  \bibinfo{volume}{2}, \bibinfo{number}{OOPSLA}, Article
  \bibinfo{articleno}{156} (\bibinfo{date}{Oct.} \bibinfo{year}{2018}),
  \bibinfo{numpages}{29}~pages.
\newblock
\urldef\tempurl%
\url{https://doi.org/10.1145/3276526}
\showDOI{\tempurl}


\bibitem[\protect\citeauthoryear{Chen, Feng, Liu, Xing, and Zhao}{Chen
  et~al\mbox{.}}{2020c}]%
        {chen2020lost}
\bibfield{author}{\bibinfo{person}{Chunyang Chen}, \bibinfo{person}{Sidong
  Feng}, \bibinfo{person}{Zhengyang Liu}, \bibinfo{person}{Zhenchang Xing},
  {and} \bibinfo{person}{Shengdong Zhao}.} \bibinfo{year}{2020}\natexlab{c}.
\newblock \showarticletitle{From lost to found: Discover missing ui design
  semantics through recovering missing tags}.
\newblock \bibinfo{journal}{\emph{Proceedings of the ACM on Human-Computer
  Interaction}} \bibinfo{volume}{4}, \bibinfo{number}{CSCW2}
  (\bibinfo{year}{2020}), \bibinfo{pages}{1--22}.
\newblock


\bibitem[\protect\citeauthoryear{Chen, Feng, Xing, Liu, Zhao, and Wang}{Chen
  et~al\mbox{.}}{2019}]%
        {galleryDC}
\bibfield{author}{\bibinfo{person}{Chunyang Chen}, \bibinfo{person}{Sidong
  Feng}, \bibinfo{person}{Zhenchang Xing}, \bibinfo{person}{Linda Liu},
  \bibinfo{person}{Shengdong Zhao}, {and} \bibinfo{person}{Jinshui Wang}.}
  \bibinfo{year}{2019}\natexlab{}.
\newblock \showarticletitle{Gallery D.C.: Design Search and Knowledge Discovery
  through Auto-Created GUI Component Gallery}.
\newblock \bibinfo{journal}{\emph{Proc. ACM Hum.-Comput. Interact.}}
  \bibinfo{volume}{3}, \bibinfo{number}{CSCW}, Article \bibinfo{articleno}{180}
  (\bibinfo{date}{Nov.} \bibinfo{year}{2019}), \bibinfo{numpages}{22}~pages.
\newblock
\urldef\tempurl%
\url{https://doi.org/10.1145/3359282}
\showDOI{\tempurl}


\bibitem[\protect\citeauthoryear{Chen, Su, Meng, Xing, and Liu}{Chen
  et~al\mbox{.}}{2018}]%
        {chunyangui2code}
\bibfield{author}{\bibinfo{person}{Chunyang Chen}, \bibinfo{person}{Ting Su},
  \bibinfo{person}{Guozhu Meng}, \bibinfo{person}{Zhenchang Xing}, {and}
  \bibinfo{person}{Yang Liu}.} \bibinfo{year}{2018}\natexlab{}.
\newblock \showarticletitle{From UI Design Image to GUI Skeleton: A Neural
  Machine Translator to Bootstrap Mobile GUI Implementation}. In
  \bibinfo{booktitle}{\emph{Proceedings of the 40th International Conference on
  Software Engineering}} (Gothenburg, Sweden) \emph{(\bibinfo{series}{ICSE
  '18})}. \bibinfo{publisher}{Association for Computing Machinery},
  \bibinfo{address}{New York, NY, USA}, \bibinfo{pages}{665–676}.
\newblock
\showISBNx{9781450356381}
\urldef\tempurl%
\url{https://doi.org/10.1145/3180155.3180240}
\showDOI{\tempurl}


\bibitem[\protect\citeauthoryear{Chen, Chen, Xing, Xia, Zhu, Grundy, and
  Wang}{Chen et~al\mbox{.}}{2020a}]%
        {Chen_uisearch}
\bibfield{author}{\bibinfo{person}{Jieshan Chen}, \bibinfo{person}{Chunyang
  Chen}, \bibinfo{person}{Zhenchang Xing}, \bibinfo{person}{Xin Xia},
  \bibinfo{person}{Liming Zhu}, \bibinfo{person}{John Grundy}, {and}
  \bibinfo{person}{Jinshui Wang}.} \bibinfo{year}{2020}\natexlab{a}.
\newblock \showarticletitle{Wireframe-based UI Design Search through Image
  Autoencoder}.
\newblock \bibinfo{journal}{\emph{ACM Transactions on Software Engineering and
  Methodology}} \bibinfo{volume}{29}, \bibinfo{number}{3} (\bibinfo{date}{Jul}
  \bibinfo{year}{2020}), \bibinfo{pages}{1–31}.
\newblock
\showISSN{1557-7392}
\urldef\tempurl%
\url{https://doi.org/10.1145/3391613}
\showDOI{\tempurl}


\bibitem[\protect\citeauthoryear{Chen, Chen, Xing, Xu, Zhut, Li, and Wang}{Chen
  et~al\mbox{.}}{2020b}]%
        {chenjieshan2020ase}
\bibfield{author}{\bibinfo{person}{Jieshan Chen}, \bibinfo{person}{Chunyang
  Chen}, \bibinfo{person}{Zhenchang Xing}, \bibinfo{person}{Xiwei Xu},
  \bibinfo{person}{Liming Zhut}, \bibinfo{person}{Guoqiang Li}, {and}
  \bibinfo{person}{Jinshui Wang}.} \bibinfo{year}{2020}\natexlab{b}.
\newblock \showarticletitle{Unblind Your Apps: Predicting Natural-Language
  Labels for Mobile GUI Components by Deep Learning}. In
  \bibinfo{booktitle}{\emph{2020 IEEE/ACM 42nd International Conference on
  Software Engineering (ICSE)}}. \bibinfo{pages}{322--334}.
\newblock


\bibitem[\protect\citeauthoryear{Chen, Swearngin, Wu, Barik, Nichols, and
  Zhang}{Chen et~al\mbox{.}}{2022a}]%
        {chen2022extracting}
\bibfield{author}{\bibinfo{person}{Jieshan Chen}, \bibinfo{person}{Amanda
  Swearngin}, \bibinfo{person}{Jason Wu}, \bibinfo{person}{Titus Barik},
  \bibinfo{person}{Jeffrey Nichols}, {and} \bibinfo{person}{Xiaoyi Zhang}.}
  \bibinfo{year}{2022}\natexlab{a}.
\newblock \showarticletitle{Extracting Replayable Interactions from Videos of
  Mobile App Usage}.
\newblock \bibinfo{journal}{\emph{arXiv preprint arXiv:2207.04165}}
  (\bibinfo{year}{2022}).
\newblock


\bibitem[\protect\citeauthoryear{Chen, Swearngin, Wu, Barik, Nichols, and
  Zhang}{Chen et~al\mbox{.}}{2022b}]%
        {chen2022towards}
\bibfield{author}{\bibinfo{person}{Jieshan Chen}, \bibinfo{person}{Amanda
  Swearngin}, \bibinfo{person}{Jason Wu}, \bibinfo{person}{Titus Barik},
  \bibinfo{person}{Jeffrey Nichols}, {and} \bibinfo{person}{Xiaoyi Zhang}.}
  \bibinfo{year}{2022}\natexlab{b}.
\newblock \showarticletitle{Towards Complete Icon Labeling in Mobile
  Applications}. In \bibinfo{booktitle}{\emph{CHI Conference on Human Factors
  in Computing Systems}}. \bibinfo{pages}{1--14}.
\newblock


\bibitem[\protect\citeauthoryear{Chen, Xie, Xing, Chen, Xu, Zhu, and Li}{Chen
  et~al\mbox{.}}{2020d}]%
        {UIED-full-fse}
\bibfield{author}{\bibinfo{person}{Jieshan Chen}, \bibinfo{person}{Mulong Xie},
  \bibinfo{person}{Zhenchang Xing}, \bibinfo{person}{Chunyang Chen},
  \bibinfo{person}{Xiwei Xu}, \bibinfo{person}{Liming Zhu}, {and}
  \bibinfo{person}{Guoqiang Li}.} \bibinfo{year}{2020}\natexlab{d}.
\newblock \showarticletitle{Object detection for graphical user interface: old
  fashioned or deep learning or a combination?}
\newblock \bibinfo{journal}{\emph{Proceedings of the 28th ACM Joint Meeting on
  European Software Engineering Conference and Symposium on the Foundations of
  Software Engineering}} (\bibinfo{date}{Nov} \bibinfo{year}{2020}).
\newblock
\showISBNx{9781450370431}
\urldef\tempurl%
\url{https://doi.org/10.1145/3368089.3409691}
\showDOI{\tempurl}


\bibitem[\protect\citeauthoryear{Degott, Borges~Jr., and Zeller}{Degott
  et~al\mbox{.}}{2019}]%
        {issta2019learninguielementinteractions}
\bibfield{author}{\bibinfo{person}{Christian Degott},
  \bibinfo{person}{Nataniel~P. Borges~Jr.}, {and} \bibinfo{person}{Andreas
  Zeller}.} \bibinfo{year}{2019}\natexlab{}.
\newblock \showarticletitle{Learning User Interface Element Interactions}. In
  \bibinfo{booktitle}{\emph{Proceedings of the 28th ACM SIGSOFT International
  Symposium on Software Testing and Analysis}} (Beijing, China)
  \emph{(\bibinfo{series}{ISSTA 2019})}. \bibinfo{publisher}{Association for
  Computing Machinery}, \bibinfo{address}{New York, NY, USA},
  \bibinfo{pages}{296–306}.
\newblock
\showISBNx{9781450362245}
\urldef\tempurl%
\url{https://doi.org/10.1145/3293882.3330569}
\showDOI{\tempurl}


\bibitem[\protect\citeauthoryear{Deka, Huang, Franzen, Hibschman, Afergan, Li,
  Nichols, and Kumar}{Deka et~al\mbox{.}}{2017}]%
        {rico}
\bibfield{author}{\bibinfo{person}{Biplab Deka}, \bibinfo{person}{Zifeng
  Huang}, \bibinfo{person}{Chad Franzen}, \bibinfo{person}{Joshua Hibschman},
  \bibinfo{person}{Daniel Afergan}, \bibinfo{person}{Yang Li},
  \bibinfo{person}{Jeffrey Nichols}, {and} \bibinfo{person}{Ranjitha Kumar}.}
  \bibinfo{year}{2017}\natexlab{}.
\newblock \showarticletitle{Rico: A Mobile App Dataset for Building Data-Driven
  Design Applications}. In \bibinfo{booktitle}{\emph{Proceedings of the 30th
  Annual Symposium on User Interface Software and Technology}}
  \emph{(\bibinfo{series}{UIST '17})}.
\newblock


\bibitem[\protect\citeauthoryear{Dribbble}{Dribbble}{[n.d.]}]%
        {dribbble}
\bibfield{author}{\bibinfo{person}{Dribbble}.}
  \bibinfo{year}{[n.d.]}\natexlab{}.
\newblock \bibinfo{title}{Discover the world's Top Designers \&amp; Creatives}.
\newblock
\newblock
\urldef\tempurl%
\url{https://dribbble.com/}
\showURL{%
\tempurl}


\bibitem[\protect\citeauthoryear{Ester, Kriegel, Sander, and Xu}{Ester
  et~al\mbox{.}}{1996}]%
        {dbscan}
\bibfield{author}{\bibinfo{person}{Martin Ester}, \bibinfo{person}{Hans-Peter
  Kriegel}, \bibinfo{person}{J\"{o}rg Sander}, {and} \bibinfo{person}{Xiaowei
  Xu}.} \bibinfo{year}{1996}\natexlab{}.
\newblock \showarticletitle{A Density-Based Algorithm for Discovering Clusters
  in Large Spatial Databases with Noise}. In
  \bibinfo{booktitle}{\emph{Proceedings of the Second International Conference
  on Knowledge Discovery and Data Mining}} (Portland, Oregon)
  \emph{(\bibinfo{series}{KDD'96})}. \bibinfo{publisher}{AAAI Press},
  \bibinfo{pages}{226–231}.
\newblock


\bibitem[\protect\citeauthoryear{Eysenck and Brysbaert}{Eysenck and
  Brysbaert}{2018}]%
        {FundamentalsofCognition}
\bibfield{author}{\bibinfo{person}{Michael~W. Eysenck} {and}
  \bibinfo{person}{Marc Brysbaert}.} \bibinfo{year}{2018}\natexlab{}.
\newblock \showarticletitle{Fundamentals of Cognition}.
\newblock  (\bibinfo{year}{2018}).
\newblock
\urldef\tempurl%
\url{https://doi.org/10.4324/9781315617633}
\showDOI{\tempurl}


\bibitem[\protect\citeauthoryear{Feng and Chen}{Feng and Chen}{2022a}]%
        {feng2022gifdroid2}
\bibfield{author}{\bibinfo{person}{Sidong Feng} {and} \bibinfo{person}{Chunyang
  Chen}.} \bibinfo{year}{2022}\natexlab{a}.
\newblock \showarticletitle{GIFdroid: An Automated Light-weight Tool for
  Replaying Visual Bug Reports}.
\newblock  (\bibinfo{year}{2022}).
\newblock


\bibitem[\protect\citeauthoryear{Feng and Chen}{Feng and Chen}{2022b}]%
        {feng2022gifdroid1}
\bibfield{author}{\bibinfo{person}{Sidong Feng} {and} \bibinfo{person}{Chunyang
  Chen}.} \bibinfo{year}{2022}\natexlab{b}.
\newblock \showarticletitle{GIFdroid: automated replay of visual bug reports
  for Android apps}. In \bibinfo{booktitle}{\emph{Proceedings of the 44th
  International Conference on Software Engineering}}.
  \bibinfo{pages}{1045--1057}.
\newblock


\bibitem[\protect\citeauthoryear{Feng, Jiang, Zhou, Zhen, and Chen}{Feng
  et~al\mbox{.}}{2022}]%
        {feng2022auto}
\bibfield{author}{\bibinfo{person}{Sidong Feng}, \bibinfo{person}{Minmin
  Jiang}, \bibinfo{person}{Tingting Zhou}, \bibinfo{person}{Yankun Zhen}, {and}
  \bibinfo{person}{Chunyang Chen}.} \bibinfo{year}{2022}\natexlab{}.
\newblock \showarticletitle{Auto-Icon+: An Automated End-to-End Code Generation
  Tool for Icon Designs in UI Development}.
\newblock \bibinfo{journal}{\emph{ACM Transactions on Interactive Intelligent
  Systems (TiiS)}} (\bibinfo{year}{2022}).
\newblock


\bibitem[\protect\citeauthoryear{Feng, Ma, Yu, Chen, Zhou, and Zhen}{Feng
  et~al\mbox{.}}{2021}]%
        {feng2021auto}
\bibfield{author}{\bibinfo{person}{Sidong Feng}, \bibinfo{person}{Suyu Ma},
  \bibinfo{person}{Jinzhong Yu}, \bibinfo{person}{Chunyang Chen},
  \bibinfo{person}{TingTing Zhou}, {and} \bibinfo{person}{Yankun Zhen}.}
  \bibinfo{year}{2021}\natexlab{}.
\newblock \showarticletitle{Auto-icon: An automated code generation tool for
  icon designs assisting in ui development}. In \bibinfo{booktitle}{\emph{26th
  International Conference on Intelligent User Interfaces}}.
  \bibinfo{pages}{59--69}.
\newblock


\bibitem[\protect\citeauthoryear{Girshick}{Girshick}{2015}]%
        {frcnn}
\bibfield{author}{\bibinfo{person}{Ross Girshick}.}
  \bibinfo{year}{2015}\natexlab{}.
\newblock \showarticletitle{Fast R-CNN}. In \bibinfo{booktitle}{\emph{The IEEE
  International Conference on Computer Vision (ICCV)}}.
\newblock


\bibitem[\protect\citeauthoryear{Guo, Li, Lou, Yang, and Liu}{Guo
  et~al\mbox{.}}{2019}]%
        {guo2019sara}
\bibfield{author}{\bibinfo{person}{Jiaqi Guo}, \bibinfo{person}{Shuyue Li},
  \bibinfo{person}{Jian-Guang Lou}, \bibinfo{person}{Zijiang Yang}, {and}
  \bibinfo{person}{Ting Liu}.} \bibinfo{year}{2019}\natexlab{}.
\newblock \showarticletitle{SARA: self-replay augmented record and replay for
  Android in industrial cases}. In \bibinfo{booktitle}{\emph{Proceedings of the
  28th acm sigsoft international symposium on software testing and analysis}}.
  \bibinfo{pages}{90--100}.
\newblock


\bibitem[\protect\citeauthoryear{Knyazev, de~Vries, Cangea, Taylor, Courville,
  and Belilovsky}{Knyazev et~al\mbox{.}}{2020}]%
        {knyazev2020graph}
\bibfield{author}{\bibinfo{person}{Boris Knyazev}, \bibinfo{person}{Harm de
  Vries}, \bibinfo{person}{Cătălina Cangea}, \bibinfo{person}{Graham~W.
  Taylor}, \bibinfo{person}{Aaron Courville}, {and} \bibinfo{person}{Eugene
  Belilovsky}.} \bibinfo{year}{2020}\natexlab{}.
\newblock \bibinfo{title}{Graph Density-Aware Losses for Novel Compositions in
  Scene Graph Generation}.
\newblock
\newblock
\showeprint[arxiv]{2005.08230}~[cs.CV]


\bibitem[\protect\citeauthoryear{Lecun, Bengio, and Hinton}{Lecun
  et~al\mbox{.}}{2015}]%
        {deeplearning_Nature}
\bibfield{author}{\bibinfo{person}{Yann Lecun}, \bibinfo{person}{Yoshua
  Bengio}, {and} \bibinfo{person}{Geoffrey Hinton}.}
  \bibinfo{year}{2015}\natexlab{}.
\newblock \showarticletitle{Deep learning}.
\newblock \bibinfo{journal}{\emph{Nature}} \bibinfo{volume}{521},
  \bibinfo{number}{7553} (\bibinfo{year}{2015}), \bibinfo{pages}{436–444}.
\newblock
\urldef\tempurl%
\url{https://doi.org/10.1038/nature14539}
\showDOI{\tempurl}


\bibitem[\protect\citeauthoryear{Li, Yang, Guo, and Chen}{Li
  et~al\mbox{.}}{2019}]%
        {ase2019humanoid}
\bibfield{author}{\bibinfo{person}{Yuanchun Li}, \bibinfo{person}{Ziyue Yang},
  \bibinfo{person}{Yao Guo}, {and} \bibinfo{person}{Xiangqun Chen}.}
  \bibinfo{year}{2019}\natexlab{}.
\newblock \showarticletitle{Humanoid: A Deep Learning-Based Approach to
  Automated Black-box Android App Testing}. In \bibinfo{booktitle}{\emph{2019
  34th IEEE/ACM International Conference on Automated Software Engineering
  (ASE)}}. \bibinfo{pages}{1070--1073}.
\newblock
\urldef\tempurl%
\url{https://doi.org/10.1109/ASE.2019.00104}
\showDOI{\tempurl}


\bibitem[\protect\citeauthoryear{Liu, Craft, Situ, Yumer, Mech, and Kumar}{Liu
  et~al\mbox{.}}{2018}]%
        {rico2}
\bibfield{author}{\bibinfo{person}{Thomas~F. Liu}, \bibinfo{person}{Mark
  Craft}, \bibinfo{person}{Jason Situ}, \bibinfo{person}{Ersin Yumer},
  \bibinfo{person}{Radomir Mech}, {and} \bibinfo{person}{Ranjitha Kumar}.}
  \bibinfo{year}{2018}\natexlab{}.
\newblock \showarticletitle{Learning Design Semantics for Mobile Apps}. In
  \bibinfo{booktitle}{\emph{The 31st Annual ACM Symposium on User Interface
  Software and Technology}} (Berlin, Germany) \emph{(\bibinfo{series}{UIST
  '18})}. \bibinfo{publisher}{ACM}, \bibinfo{address}{New York, NY, USA},
  \bibinfo{pages}{569--579}.
\newblock
\showISBNx{978-1-4503-5948-1}
\urldef\tempurl%
\url{https://doi.org/10.1145/3242587.3242650}
\showDOI{\tempurl}


\bibitem[\protect\citeauthoryear{Moran, Bernal-Cárdenas, Curcio, Bonett, and
  Poshyvanyk}{Moran et~al\mbox{.}}{2020}]%
        {redraw}
\bibfield{author}{\bibinfo{person}{Kevin Moran}, \bibinfo{person}{Carlos
  Bernal-Cárdenas}, \bibinfo{person}{Michael Curcio}, \bibinfo{person}{Richard
  Bonett}, {and} \bibinfo{person}{Denys Poshyvanyk}.}
  \bibinfo{year}{2020}\natexlab{}.
\newblock \showarticletitle{Machine Learning-Based Prototyping of Graphical
  User Interfaces for Mobile Apps}.
\newblock \bibinfo{journal}{\emph{IEEE Transactions on Software Engineering}}
  \bibinfo{volume}{46}, \bibinfo{number}{2} (\bibinfo{year}{2020}),
  \bibinfo{pages}{196--221}.
\newblock
\urldef\tempurl%
\url{https://doi.org/10.1109/TSE.2018.2844788}
\showDOI{\tempurl}


\bibitem[\protect\citeauthoryear{Nguyen and Csallner}{Nguyen and
  Csallner}{2015}]%
        {remaui}
\bibfield{author}{\bibinfo{person}{Tuan~Anh Nguyen} {and}
  \bibinfo{person}{Christoph Csallner}.} \bibinfo{year}{2015}\natexlab{}.
\newblock \showarticletitle{Reverse Engineering Mobile Application User
  Interfaces with REMAUI}. In \bibinfo{booktitle}{\emph{Proceedings of the 30th
  IEEE/ACM International Conference on Automated Software Engineering}}
  (Lincoln, Nebraska) \emph{(\bibinfo{series}{ASE '15})}.
  \bibinfo{publisher}{IEEE Press}, \bibinfo{pages}{248–259}.
\newblock
\showISBNx{9781509000241}
\urldef\tempurl%
\url{https://doi.org/10.1109/ASE.2015.32}
\showDOI{\tempurl}


\bibitem[\protect\citeauthoryear{Qian, Shang, Yan, Wang, and Chen}{Qian
  et~al\mbox{.}}{2020}]%
        {roscript}
\bibfield{author}{\bibinfo{person}{Ju Qian}, \bibinfo{person}{Zhengyu Shang},
  \bibinfo{person}{Shuoyan Yan}, \bibinfo{person}{Yan Wang}, {and}
  \bibinfo{person}{Lin Chen}.} \bibinfo{year}{2020}\natexlab{}.
\newblock \showarticletitle{RoScript: A Visual Script Driven Truly
  Non-Intrusive Robotic Testing System for Touch Screen Applications}. In
  \bibinfo{booktitle}{\emph{Proceedings of the ACM/IEEE 42nd International
  Conference on Software Engineering}} (Seoul, South Korea)
  \emph{(\bibinfo{series}{ICSE '20})}. \bibinfo{publisher}{Association for
  Computing Machinery}, \bibinfo{address}{New York, NY, USA},
  \bibinfo{pages}{297–308}.
\newblock
\showISBNx{9781450371216}
\urldef\tempurl%
\url{https://doi.org/10.1145/3377811.3380431}
\showDOI{\tempurl}


\bibitem[\protect\citeauthoryear{Redmon and Farhadi}{Redmon and
  Farhadi}{2018}]%
        {yolo}
\bibfield{author}{\bibinfo{person}{Joseph Redmon} {and} \bibinfo{person}{Ali
  Farhadi}.} \bibinfo{year}{2018}\natexlab{}.
\newblock \showarticletitle{YOLOv3: An Incremental Improvement}.
\newblock \bibinfo{journal}{\emph{CoRR}}  \bibinfo{volume}{abs/1804.02767}
  (\bibinfo{year}{2018}).
\newblock
\showeprint[arxiv]{1804.02767}
\urldef\tempurl%
\url{http://arxiv.org/abs/1804.02767}
\showURL{%
\tempurl}


\bibitem[\protect\citeauthoryear{Rutledge}{Rutledge}{2009}]%
        {andyrutledge_3laws}
\bibfield{author}{\bibinfo{person}{Andy Rutledge}.}
  \bibinfo{year}{2009}\natexlab{}.
\newblock \bibinfo{title}{Gestalt Principles of Perception - 3: Proximity,
  Uniform Connectedness, and Good Continuation}.
\newblock
\newblock
\urldef\tempurl%
\url{http://andyrutledge.com/gestalt-principles-3.html}
\showURL{%
\tempurl}


\bibitem[\protect\citeauthoryear{S1T2}{S1T2}{2021}]%
        {crap}
\bibfield{author}{\bibinfo{person}{S1T2}.} \bibinfo{year}{2021}\natexlab{}.
\newblock \bibinfo{title}{Apply crap to design: S1T2 blog}.
\newblock
\newblock
\urldef\tempurl%
\url{https://s1t2.com/blog/step-1-generously-apply-crap-to-design}
\showURL{%
\tempurl}


\bibitem[\protect\citeauthoryear{Sahin, Aliyeva, Mathavan, Coskun, and
  Egele}{Sahin et~al\mbox{.}}{2019}]%
        {sahin2019randr}
\bibfield{author}{\bibinfo{person}{Onur Sahin}, \bibinfo{person}{Assel
  Aliyeva}, \bibinfo{person}{Hariharan Mathavan}, \bibinfo{person}{Ayse
  Coskun}, {and} \bibinfo{person}{Manuel Egele}.}
  \bibinfo{year}{2019}\natexlab{}.
\newblock \showarticletitle{Randr: Record and replay for android applications
  via targeted runtime instrumentation}. In \bibinfo{booktitle}{\emph{2019 34th
  IEEE/ACM International Conference on Automated Software Engineering (ASE)}}.
  IEEE, \bibinfo{pages}{128--138}.
\newblock


\bibitem[\protect\citeauthoryear{Sketch}{Sketch}{[n.d.]}]%
        {sketch}
\bibfield{author}{\bibinfo{person}{Sketch}.} \bibinfo{year}{[n.d.]}\natexlab{}.
\newblock
\newblock
\urldef\tempurl%
\url{https://www.sketch.com/}
\showURL{%
\tempurl}


\bibitem[\protect\citeauthoryear{Sternberg and Robert}{Sternberg and
  Robert}{2003}]%
        {CognitivePsychology}
\bibfield{author}{\bibinfo{person}{Sternberg} {and} \bibinfo{person}{Robert}.}
  \bibinfo{year}{2003}\natexlab{}.
\newblock \bibinfo{booktitle}{\emph{Cognitive Psychology Third Edition}}.
\newblock \bibinfo{publisher}{Thomson Wadsworth}.
\newblock


\bibitem[\protect\citeauthoryear{Stevenson}{Stevenson}{[n.d.]}]%
        {TheGestaltApproachtoChange}
\bibfield{author}{\bibinfo{person}{Herb Stevenson}.}
  \bibinfo{year}{[n.d.]}\natexlab{}.
\newblock \bibinfo{title}{Emergence: The Gestalt Approach to Change}.
\newblock
\newblock
\urldef\tempurl%
\url{http://www.clevelandconsultinggroup.com/articles/emergence-gestalt-approach-to-change.php}
\showURL{%
\tempurl}


\bibitem[\protect\citeauthoryear{Tesseract-Ocr}{Tesseract-Ocr}{[n.d.]}]%
        {tesseract-ocr}
\bibfield{author}{\bibinfo{person}{Tesseract-Ocr}.}
  \bibinfo{year}{[n.d.]}\natexlab{}.
\newblock \bibinfo{title}{tesseract-ocr/tesseract: Tesseract Open Source OCR
  Engine (main repository)}.
\newblock
\newblock
\urldef\tempurl%
\url{https://github.com/tesseract-ocr/tesseract}
\showURL{%
\tempurl}


\bibitem[\protect\citeauthoryear{Thalion}{Thalion}{2020}]%
        {gelstal_for_ui_design}
\bibfield{author}{\bibinfo{person}{Thalion}.} \bibinfo{year}{2020}\natexlab{}.
\newblock \bibinfo{title}{Ui Design in practice: Gestalt principles}.
\newblock
\newblock
\urldef\tempurl%
\url{https://uxmisfit.com/2019/04/23/ui-design-in-practice-gestalt-principles/}
\showURL{%
\tempurl}


\bibitem[\protect\citeauthoryear{Uijlings, van~de Sande, Gevers, and
  Smeulders}{Uijlings et~al\mbox{.}}{2013}]%
        {selectivesearch}
\bibfield{author}{\bibinfo{person}{J.~R.~R. Uijlings},
  \bibinfo{person}{K.~E.~A. van~de Sande}, \bibinfo{person}{T. Gevers}, {and}
  \bibinfo{person}{A.~W.~M. Smeulders}.} \bibinfo{year}{2013}\natexlab{}.
\newblock \showarticletitle{Selective Search for Object Recognition}.
\newblock \bibinfo{journal}{\emph{International Journal of Computer Vision}}
  \bibinfo{volume}{104}, \bibinfo{number}{2} (\bibinfo{date}{01 Sep}
  \bibinfo{year}{2013}), \bibinfo{pages}{154--171}.
\newblock
\showISSN{1573-1405}
\urldef\tempurl%
\url{https://doi.org/10.1007/s11263-013-0620-5}
\showDOI{\tempurl}


\bibitem[\protect\citeauthoryear{UserTesting}{UserTesting}{2019}]%
        {usertesting_7GP}
\bibfield{author}{\bibinfo{person}{UserTesting}.}
  \bibinfo{year}{2019}\natexlab{}.
\newblock \bibinfo{title}{7 Gestalt Principles of Visual Perception: Cognitive
  Psychology for UX: UserTesting Blog}.
\newblock
\newblock
\urldef\tempurl%
\url{https://www.usertesting.com/blog/gestalt-principles#proximity}
\showURL{%
\tempurl}


\bibitem[\protect\citeauthoryear{Vinyals, Toshev, Bengio, and Erhan}{Vinyals
  et~al\mbox{.}}{2015}]%
        {imgcaption1}
\bibfield{author}{\bibinfo{person}{Oriol Vinyals}, \bibinfo{person}{Alexander
  Toshev}, \bibinfo{person}{Samy Bengio}, {and} \bibinfo{person}{Dumitru
  Erhan}.} \bibinfo{year}{2015}\natexlab{}.
\newblock \bibinfo{title}{Show and Tell: A Neural Image Caption Generator}.
\newblock
\newblock
\showeprint[arxiv]{1411.4555}~[cs.CV]


\bibitem[\protect\citeauthoryear{White, Fraser, and Brown}{White
  et~al\mbox{.}}{2019}]%
        {whiteyolo}
\bibfield{author}{\bibinfo{person}{Thomas~D. White}, \bibinfo{person}{Gordon
  Fraser}, {and} \bibinfo{person}{Guy~J. Brown}.}
  \bibinfo{year}{2019}\natexlab{}.
\newblock \showarticletitle{Improving Random GUI Testing with Image-based
  Widget Detection}. In \bibinfo{booktitle}{\emph{Proceedings of the 28th ACM
  SIGSOFT International Symposium on Software Testing and Analysis}} (Beijing,
  China) \emph{(\bibinfo{series}{ISSTA 2019})}. \bibinfo{publisher}{ACM},
  \bibinfo{address}{New York, NY, USA}, \bibinfo{pages}{307--317}.
\newblock
\showISBNx{978-1-4503-6224-5}
\urldef\tempurl%
\url{https://doi.org/10.1145/3293882.3330551}
\showDOI{\tempurl}


\bibitem[\protect\citeauthoryear{Xie, Feng, Xing, Chen, and Chen}{Xie
  et~al\mbox{.}}{2020}]%
        {xie2020uied}
\bibfield{author}{\bibinfo{person}{Mulong Xie}, \bibinfo{person}{Sidong Feng},
  \bibinfo{person}{Zhenchang Xing}, \bibinfo{person}{Jieshan Chen}, {and}
  \bibinfo{person}{Chunyang Chen}.} \bibinfo{year}{2020}\natexlab{}.
\newblock \showarticletitle{UIED: a hybrid tool for GUI element detection}. In
  \bibinfo{booktitle}{\emph{Proceedings of the 28th ACM Joint Meeting on
  European Software Engineering Conference and Symposium on the Foundations of
  Software Engineering}}. \bibinfo{pages}{1655--1659}.
\newblock


\bibitem[\protect\citeauthoryear{Yandrapally, Stocco, and Mesbah}{Yandrapally
  et~al\mbox{.}}{2020}]%
        {nearduplicate1}
\bibfield{author}{\bibinfo{person}{Rahulkrishna Yandrapally},
  \bibinfo{person}{Andrea Stocco}, {and} \bibinfo{person}{Ali Mesbah}.}
  \bibinfo{year}{2020}\natexlab{}.
\newblock \showarticletitle{Near-Duplicate Detection in Web App Model
  Inference}. In \bibinfo{booktitle}{\emph{Proceedings of the ACM/IEEE 42nd
  International Conference on Software Engineering}} (Seoul, South Korea)
  \emph{(\bibinfo{series}{ICSE '20})}. \bibinfo{publisher}{Association for
  Computing Machinery}, \bibinfo{address}{New York, NY, USA},
  \bibinfo{pages}{186–197}.
\newblock
\showISBNx{9781450371216}
\urldef\tempurl%
\url{https://doi.org/10.1145/3377811.3380416}
\showDOI{\tempurl}


\bibitem[\protect\citeauthoryear{Yang, Lu, Lee, Batra, and Parikh}{Yang
  et~al\mbox{.}}{2018}]%
        {yang2018graph}
\bibfield{author}{\bibinfo{person}{Jianwei Yang}, \bibinfo{person}{Jiasen Lu},
  \bibinfo{person}{Stefan Lee}, \bibinfo{person}{Dhruv Batra}, {and}
  \bibinfo{person}{Devi Parikh}.} \bibinfo{year}{2018}\natexlab{}.
\newblock \bibinfo{title}{Graph R-CNN for Scene Graph Generation}.
\newblock
\newblock
\showeprint[arxiv]{1808.00191}~[cs.CV]


\bibitem[\protect\citeauthoryear{YazdaniBanafsheDaragh}{YazdaniBanafsheDaragh}{[n.d.]}]%
        {deepgui}
\bibfield{author}{\bibinfo{person}{YazdaniBanafsheDaragh}.}
  \bibinfo{year}{[n.d.]}\natexlab{}.
\newblock \showarticletitle{Deep-GUI: Towards Platform-Independent UI Input
  Generation with Deep Reinforcement Learning}.
\newblock \bibinfo{journal}{\emph{UC Irvine}} (\bibinfo{year}{[n.\,d.]}).
\newblock
\urldef\tempurl%
\url{https://escholarship.org/uc/item/3kv1n3qk}
\showURL{%
\tempurl}


\bibitem[\protect\citeauthoryear{Zellers, Yatskar, Thomson, and Choi}{Zellers
  et~al\mbox{.}}{2018}]%
        {zellers2018neural}
\bibfield{author}{\bibinfo{person}{Rowan Zellers}, \bibinfo{person}{Mark
  Yatskar}, \bibinfo{person}{Sam Thomson}, {and} \bibinfo{person}{Yejin Choi}.}
  \bibinfo{year}{2018}\natexlab{}.
\newblock \bibinfo{title}{Neural Motifs: Scene Graph Parsing with Global
  Context}.
\newblock
\newblock
\showeprint[arxiv]{1711.06640}~[cs.CV]


\bibitem[\protect\citeauthoryear{Zhang, de~Greef, Swearngin, White, Murray, Yu,
  Shan, Nichols, Wu, Fleizach, Everitt, and Bigham}{Zhang
  et~al\mbox{.}}{2021}]%
        {screenrecognition}
\bibfield{author}{\bibinfo{person}{Xiaoyi Zhang}, \bibinfo{person}{Lilian de
  Greef}, \bibinfo{person}{Amanda Swearngin}, \bibinfo{person}{Samuel White},
  \bibinfo{person}{Kyle Murray}, \bibinfo{person}{Lisa Yu}, \bibinfo{person}{Qi
  Shan}, \bibinfo{person}{Jeffrey Nichols}, \bibinfo{person}{Jason Wu},
  \bibinfo{person}{Chris Fleizach}, \bibinfo{person}{Aaron Everitt}, {and}
  \bibinfo{person}{Jeffrey~P. Bigham}.} \bibinfo{year}{2021}\natexlab{}.
\newblock \bibinfo{title}{Screen Recognition: Creating Accessibility Metadata
  for Mobile Applications from Pixels}.
\newblock
\newblock
\showeprint[arxiv]{2101.04893}~[cs.HC]


\bibitem[\protect\citeauthoryear{Zheng, Hu, and Ma}{Zheng
  et~al\mbox{.}}{2019}]%
        {faceoff}
\bibfield{author}{\bibinfo{person}{Shuyu Zheng}, \bibinfo{person}{Ziniu Hu},
  {and} \bibinfo{person}{Yun Ma}.} \bibinfo{year}{2019}\natexlab{}.
\newblock \showarticletitle{FaceOff: Assisting the Manifestation Design of Web
  Graphical User Interface}. In \bibinfo{booktitle}{\emph{Proceedings of the
  Twelfth ACM International Conference on Web Search and Data Mining}}
  (Melbourne VIC, Australia) \emph{(\bibinfo{series}{WSDM '19})}.
  \bibinfo{publisher}{Association for Computing Machinery},
  \bibinfo{address}{New York, NY, USA}, \bibinfo{pages}{774–777}.
\newblock
\showISBNx{9781450359405}
\urldef\tempurl%
\url{https://doi.org/10.1145/3289600.3290610}
\showDOI{\tempurl}


\bibitem[\protect\citeauthoryear{Zhou, Yao, Wen, Wang, Zhou, He, and
  Liang}{Zhou et~al\mbox{.}}{2017}]%
        {east}
\bibfield{author}{\bibinfo{person}{Xinyu Zhou}, \bibinfo{person}{Cong Yao},
  \bibinfo{person}{He Wen}, \bibinfo{person}{Yuzhi Wang},
  \bibinfo{person}{Shuchang Zhou}, \bibinfo{person}{Weiran He}, {and}
  \bibinfo{person}{Jiajun Liang}.} \bibinfo{year}{2017}\natexlab{}.
\newblock \showarticletitle{EAST: an efficient and accurate scene text
  detector}. \bibinfo{pages}{5551--5560}.
\newblock


\end{thebibliography}

\end{document}